\documentclass[%
 reprint,
 amsmath,amssymb,
 aps,
]{revtex4-1}

\usepackage{graphicx}% Include figure files
\usepackage{dcolumn}% Align table columns on decimal point
\usepackage{bm}% bold math

\newcommand{\btwo}{\ensuremath{B_{2}}}

\newcommand{\bA}{\ensuremath{B_{A}}}

\newcommand{\tritium}{\ensuremath{{}^{3}\mathrm{H}}}
\newcommand{\hethree}{\ensuremath{{}^{3}\mathrm{He}}}
\newcommand{\hefour}{\ensuremath{{}^{4}\mathrm{He}}}
\newcommand{\hthreelambda}{\ensuremath{{}^{3}_{\Lambda}\mathrm{H}}}

\newcommand{\antitritium}{\ensuremath{{}^{3}\overline{\mathrm{H}}}}
\newcommand{\antihethree}{\ensuremath{{}^{3}\overline{\mathrm{He}}}}

\newcommand{\dradius}{\ensuremath{r_{d}}}
\newcommand{\rperp}{\ensuremath{R_{\perp}}}
\newcommand{\rpar}{\ensuremath{R_{\|}}}
\newcommand{\rmsradius}{\ensuremath{\lambda_{A}}}

\newcommand{\GeVc}{GeV/$c$}

\newcommand{\PbPb}         {\mbox{Pb--Pb}}
\newcommand{\pPb}          {\mbox{p--Pb}}

\newcommand{\avdNdeta}       {\ensuremath{\left<\mathrm{d}N_\mathrm{ch}/\mathrm{d}\eta\right>}}

\newcommand{\pt}           {\ensuremath{p_{\mathrm{T}}}}

\begin{document}

\preprint{APS/123-QED}

\title{Testing production scenarios for (anti-)(hyper-)nuclei and exotica at energies available at the CERN Large Hadron Collider}

\author{Francesca Bellini}
\email{francesca.bellini@cern.ch}
\author{Alexander P. Kalweit}
\email{alexander.kalweit@cern.ch}
\affiliation{European Organisation for Nuclear Research (CERN) \\1211 Geneva, Switzerland}

\date{\today}

\begin{abstract}
We present a detailed comparison of coalescence and thermal-statistical models for the production of (anti-)(hyper-)nuclei in high-energy collisions. For the first time, such a study is carried out as a function of the size of the object relative to the size of the particle emitting source. Our study reveals large differences between the two scenarios for the production of objects with extended wave-functions. While both models give similar predictions and show similar agreement with experimental data for (anti-)deuterons and (anti-)\hethree\ nuclei, they largely differ in their description of (anti-)hyper-triton production.
We propose to address experimentally the comparison of the production models by measuring the coalescence parameter systematically for different (anti-)(hyper-)nuclei in different collision systems and differentially in multiplicity. 
Such measurements are feasible with the current and upgraded Large Hadron Collider experiments. 
Our findings highlight the unique potential of ultra-relativistic heavy-ion collisions as a laboratory to clarify the internal structure of exotic QCD objects and can serve as a basis for more refined calculations in the future.
\end{abstract}

%\pacs{Valid PACS appear here}
%\keywords{(Anti-)nuclei, hyper-triton, coalescence, thermal model}

\maketitle

\section{Introduction} 
Nuclei and hyper-nuclei are special objects with respect to non-composite hadrons (pions, protons, etc.), because their size is comparable to a fraction or the whole system created in high-energy proton-proton (pp), proton-nucleus (pA) and nucleus-nucleus (AA) collisions~\cite{Adam:2015vna}.  
Their size is typically defined as the rms of their (charge) wave-function, corresponding to about 2 fm for light (anti-)nuclei as obtained from electron scattering experiments. 
For the hyper-triton, theoretical calculations indicate an rms of the wave-function of about 5 fm \cite{Nemura:1999qp}, significantly larger than that of non-strange nuclei with mass number $A = 3$ and driven by the average separation of the $\Lambda$ relative to the other two nucleons. 
This difference in the wave-functions results in dramatic consequences for the production scenarios, as discussed in the following.
The properties of the objects under study here are summarised in Table \ref{tab:nucleusradii}.

For about 60 years, coalescence models have been used to describe the formation of composite objects \cite{Butler:1963, Kapusta:1980, Sato:1981ez, Nagle:1996vp, Scheibl:1998tk, Cho:2017dcy, Blum:2017qnn, Bazak:2018hgl, Zhao:2018lyf}.
Surprisingly, thermal-statistical models have been successful in describing the production of light \mbox{(anti-)}(hyper-)nuclei across a wide range of energies in AA collisions \cite{Andronic:2010qu,Andronic:2017}. 
In this approach, particles are produced from a fireball in thermal equilibrium with temperatures of $T_{chem} \approx$ 156 MeV.
Particle abundances are fixed at chemical freeze-out, when inelastic collisions cease. Further elastic and pseudo-elastic collisions occur among the components of the expanding fireball, which affect the spectral shapes and the measurable yields of short-lived (strongly decaying) hadronic resonances. 
Once the mean free path for elastic collisions is larger than the system size, the fireball freezes-out kinetically at $T_{kin} \approx$ 90~MeV~\cite{Abelev:2013vea}. 
In such a dense and hot environment, composite objects with binding energies that are low with respect to the temperature of the system, appear as ``fragile''. 
For instance, the binding energy of the deuteron is $B_{E, d}$ = 2.2~MeV~$\ll~T_{chem},~T_{kin}$.
The cross-section for pion-induced deuteron breakup is significantly larger than the typical (pseudo-)elastic cross-sections for the re-scattering of hadronic resonance decay products \cite{Garcilazo:1982yc, Bass:1998ca, Schukraft:2017nbn, Oliinychenko:2018ugs}. 
Similarly, the elastic cross-section driving deuteron spectra to kinetic equilibration in central heavy-ion collisions \cite{Acharya:2017dmc} is smaller than the breakup cross-section \cite{Garcilazo:1982yc, Bass:1998ca, Schukraft:2017nbn, Oliinychenko:2018ugs}   
(Anti-)nuclei produced at chemical freeze-out are not expected to survive the hadronic phase, yet their measured production is consistent with statistical-thermal model predictions and a non-zero elliptic flow is observed \cite{Acharya:2017dmc, Puccio:2019oyd}. 
Several solutions have been proposed to solve this ``(anti-)nuclei puzzle'': (a.) a sudden freeze-out at the QGP-hadron phase boundary \cite{Castorina:2019pnb}, (b.) the thermal production of these objects as compact quark bags \cite{Andronic:2017}, (c.) the continuous interplay of breakup and formation reactions resulting in the coincidence of thermal and kinetic equilibration \cite{Oliinychenko:2018ugs, Xu:2018jff}, 
and (d.) the coincidence of coalescence and thermal production \cite{Scheibl:1998tk, HeinzTorino}.
Data from rescattering of short-lived hadronic resonances suggest a long-lasting hadronic phase \cite{Abelev:2014uua}, thus strongly disfavouring hypothesis (a.). 
Hypothesis (b.) cannot presently be tested beyond the agreement of measured (anti-)nuclei yields with statistical-thermal model predictions. 
Calculations for hypothesis (c.) are currently available only for deuterons. Hypothesis (d.) is scrutinised in this paper.
To this purpose, we propose a new method to compare models that also allows for a direct comparison with LHC data. 
\begin{table*}[htb]
\centering
\begin{ruledtabular}
\begin{tabular}{cccccccc}\\[-2ex]
Mass number & Nucleus &  Composition  & Binding energy (MeV)   &  Spin & \rmsradius$^{meas}$~(fm) &  $r_{A}$ (fm) & Refs. 
\\[0.5ex] \hline \\[-2ex]
      A = 2  & d                                    & pn                   &   2.224575 (9)     &     1   & 2.1413 $\pm$ 0.0025      &  3.2    &   \cite{VanDerLeun:1982bhg,Mohr:2015ccw}     \\[0.5ex]  \hline \\[-2ex]
      A = 3  & \tritium 	                  & pnn                       &    8.4817986 (20) & 1/2   &  1.755  $\pm$ 0.086        &  2.15   &   \cite{Purcell:2015gtm}           \\
                & \hethree                         & ppn                 &   7.7180428  (23) & 1/2   & 1.959 $\pm$  0.030         &   2.48  &   \cite{Purcell:2015gtm} \\
                & \hthreelambda               & p$\Lambda$n &    0.13 $\pm$ 0.05  \footnote{For the hypertriton, we report here the separation energy of the $\Lambda$ from the other two nucleons.} & 1/2  &  4.9 --  10.0                     &  6.8 -- 14.1 & \cite{Davis:2005mb,Nemura:1999qp} \\[0.5ex]                                    
\end{tabular}
\end{ruledtabular}
\caption{Properties of nuclei and hyper-nuclei with mass number $A \leq 3$. The nucleus size is given in terms of the (charge) rms radius of the wave-function, \rmsradius. The size parameter of the wave-function of the harmonic oscillator potential,  $r_{A}$, is chosen to approximately reproduce the measured/expected rms,  \rmsradius$^{meas}$~(fm). The proton rms charge radius $\lambda_{p} = 0.879(8)$ fm \cite{bernauer10} is subtracted from $\rmsradius^{meas}$  according to $\rmsradius = \sqrt{(\rmsradius^{meas})^2 - {\lambda_{p}^2}}$ to account for the finite extension of the constituents. We assume $\lambda_{\Lambda}\approx \lambda_{n}\approx \lambda_{p}$.}
\label{tab:nucleusradii}
\end{table*}

\indent For the first time, LHC data allow for the study of \mbox{(anti-)(hyper-)nuclei} production as a function of the system- and object-size. 
A quantitative comparison of the production scenarios has been proposed in \cite{Mrowczynski:2016xqm}, resulting in the idea of studying the production rates of nuclei with similar masses but very different internal structures, as \hefour~and ${}^{4}\mathrm{Li}$ \cite{Bazak:2018hgl}. 
As ${}^{4}\mathrm{Li}$ is not stable with respect to strong decay, its measurement is experimentally difficult and probably less constraining than the systematic measurement of the (hyper-)nuclei coalescence parameters proposed here.

\section{Coalescence approach} \label{sec:coalescence}
In the coalescence picture, nucleons produced in the collision coalesce into nuclei if they are close in space and have similar velocities \cite{Butler:1963,Kapusta:1980}. 
For a nucleus with mass number $A = Z + N$, the coalescence probability is quantified by the coalescence parameter $B_{A}$.
Considering that at LHC energies the number of produced protons and neutrons at midrapidity as well as their momentum distributions are expected to be equal, $B_{A}$ is defined as

\begin{equation}
E_{A}\frac{\mathrm{d}^{3}N_{A}}{\mathrm{d}p_{A}^{3}}=B_{A}{\left(E_{\mathrm{p,n}}\frac{\mathrm{d}^{3}N_{\mathrm{p,n}}}{\mathrm{d}p_{\mathrm{p,n}}^{3}}\right)^{A}}\left\vert_{\vec{p}_{\mathrm{p}}=\vec{p}_{\mathrm{n}}=\frac{\vec{p}_{A}}{A}} \right.,
\label{eq:BA}
\end{equation}
where $p_{\mathrm{p,n}}$ are the proton and neutron momenta and $E_{p,n}$ their energies.
The LHC is particularly suited for the production of anti-nuclei, since the number of baryons and anti-baryons is equal at midrapidity \cite{Abbas:2013rua}. Consequently, the anti-particle to particle ratio for \mbox{(hyper-)nuclei} is consistent with unity \cite{ALICE:nucleipp2017, Acharya:2019rgc, anielski-HQ14, Acharya:2017dmc, Adam:2015yta}.
In a simple coalescence approach, \bA~is expected to be independent of momentum and of the object size relative to the volume of particle emission (hereafter referred to as ``source size'').
While this picture is found to be approximately valid in pp and \pPb~collisions~\cite{ALICE:nucleipp2017, Acharya:2019rgc, anielski-HQ14}, it breaks down in \PbPb~collisions, which exhibit a strong decrease in $B_{A}$ with centrality \cite{ALICE:deuteronppPbPb2015}. 
The elliptic flow of deuterons cannot be explained by simple coalescence \cite{Acharya:2017dmc}. 

More advanced coalescence models \cite{Sato:1981ez, Nagle:1996vp, Scheibl:1998tk} take into account the source size, as the coalescence probability naturally decreases for nucleons with similar momenta that are produced far apart in configuration space. We rely on the formalism proposed in~\cite{Scheibl:1998tk}.
As coalescence is a quantum-mechanical process, the classical definition of phase space is replaced by the Wigner formalism. The production probability of a nucleon cluster is given by the overlap of the Wigner function with the phase-space distributions of the constituents.
The wave-functions of the (hyper-)nuclei are approximated by the ground-state wave-functions of an isotropic spherical harmonic oscillator as in~\cite{Scheibl:1998tk} with one single characteristic-size parameter, $r_{A}$. 
For the deuteron wave-function $\varphi_{d}(\vec{r}) $, one obtains

\begin{equation}\label{eq:WaveHeinz}
 \varphi_{d}(\vec{r}) = (\pi r_{d}^2)^{-3/4} \exp\left(-{r^2 \over 2 r_{d}^2}\right) \;.
\end{equation}

\noindent For nuclei with $A > 2$, analogous forms exist. The relation between $r_{A}$ and the rms of the wave-function was derived in~\cite{Shebeko:2006ud} as

\begin{equation}
 \rmsradius^2 = {3 \over 2}{A - 1 \over A} {r_{A}^2 \over 2}
\end{equation} 

\noindent for point-like constituents. 
We follow the Gaussian ansatz to obtain fully analytic solutions.
In Tab.~\ref{tab:nucleusradii}, we list the measured rms of the wave-function, $\rmsradius^{meas}$, and the $r_{A}$ parameter derived from these relations. 
We encourage future more rigorous numerical studies that address the calculation of coalescence probabilities with more realistic wave-functions, e.g. the Hulthen parameterisation for deuterons \cite{Nagle:1996vp} or a $\Lambda$-deuteron parameterisation for hyper-tritons as done in~\cite{Zhang:2018euf, Sun:2018mqq}. 

\begin{figure}[hb]
\begin{center}
\includegraphics[width=\columnwidth]{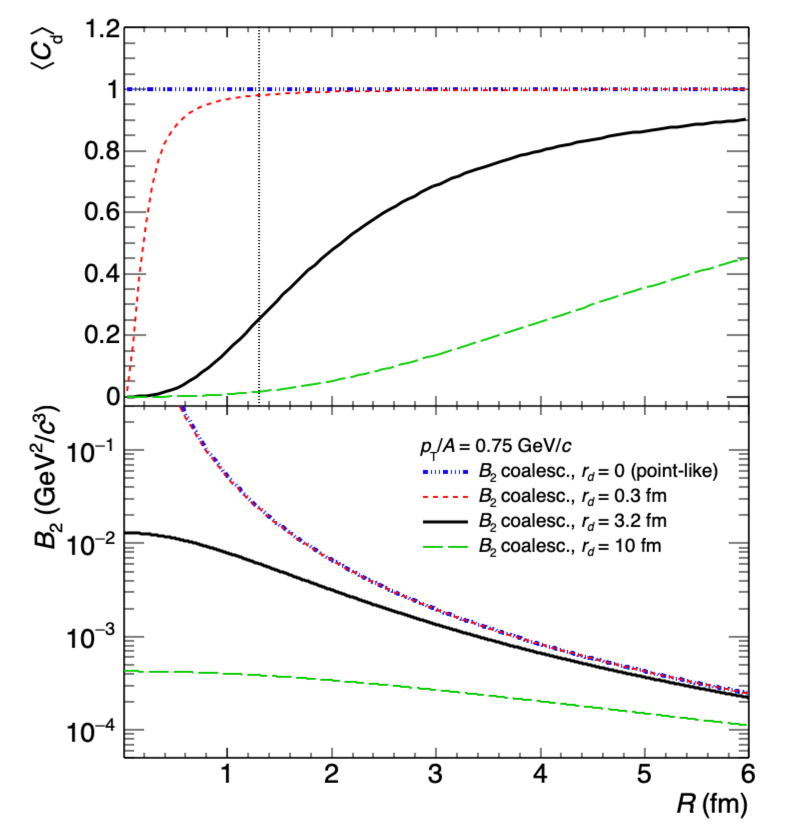}
\caption{{Quantum-mechanical correction factor $\langle C_{d} \rangle$ (top, see Eq.~\ref{eq:Cdapprox}) and coalescence parameter \btwo~for the deuteron (bottom, see Eq.~\ref{eq:B2approx}) as a function of the radius of the source $R$, calculated assuming a size parameter for the deuteron $\dradius = 0, 0.3, 3.2$ and $10$ fm. The inflection point of $\langle C_{d} \rangle$ corresponds to $R = \dradius/\sqrt{6}$ and is indicated in the left panel by the dotted vertical line for $\dradius = 3.2$ fm.}}
\label{fig:radiusDependence}
\end{center}
\end{figure}

The quantum-mechanical nature of the coalescence products is explicitly accounted for by means of an average correction factor, $\langle C_{A} \rangle$. For deuterons, $\langle C_{d} \rangle$ has been approximated as 
\begin{equation}
\langle C_{d} \rangle \approx \frac{1}{\left[1+ \left(\frac{\dradius}{2R_{\perp}(m_{T})}\right)^{2}\right] \sqrt{1+ \left(\frac{\dradius}{2R_{\|}(m_{T})}\right)^{2}}}
\label{eq:Cd}
\end{equation}
where \dradius~is the size parameter, \rperp~and \rpar~are the lengths of homogeneity of the coalescence volume and $m_{T}$ is the transverse mass of the coalescing nucleons.
The nucleus size enters the calculation of \btwo~via $\langle C_{d} \rangle$, as well as the homogeneity volume $\rperp^{2}\rpar$, according to the relation~\cite{Scheibl:1998tk}
\begin{equation}
\btwo = \frac{3\pi^{3/2} \langle C_{d} \rangle}{2m_{T}\rperp^{2}(m_{T})\rpar(m_{T})} \;\; .
\label{eq:B2cd}
\end{equation}
The coalescence parameter decreases with increasing volume, as expected. 
$\langle C_{d} \rangle$ introduces a length scale defined by the deuteron size relative to the source size. 
If we assume that  $\rperp \approx \rpar \approx R$, Eqs. \ref{eq:Cd} and \ref{eq:B2cd} simplify to 
\begin{equation}
\langle C_{d} \rangle \approx \left[1+ \left(\frac{\dradius}{2R(m_{T})}\right)^{2}\right]^{-3/2}
\label{eq:Cdapprox}
\end{equation}
and
\begin{equation}
\btwo = \frac{3\pi^{3/2} \langle C_{d} \rangle}{2m_{T}R^{3}(m_{T})}.
\label{eq:B2approx}
\end{equation}
Figure \ref{fig:radiusDependence} shows the source radius ($R$) dependence of $\langle C_{d} \rangle$ and \btwo, calculated assuming (a.) $\dradius = 0$, (b.) $\dradius = 0.3$ fm, (c.) the actual value $\dradius = 3.2$ fm~\cite{Mohr:2015ccw}, (d.) a larger, unrealistic $\dradius = 10$ fm. 
As shown in Fig. \ref{fig:radiusDependence}, $\langle C_{d} \rangle$ suppresses significantly the production of objects with a radius larger than the source. 

Following the discussion in~\cite{Scheibl:1998tk, Blum:2017qnn}, Eq.~\ref{eq:Cd} may be generalised as 
\begin{equation}
\langle C_{A} \rangle = \prod_{i=1,2,3} \left(1 + \frac{r^2}{4R_{i}^2} \right)^{-\frac{1}{2}(A-1)} \;\; .
\label{eq:CA_general}
\end{equation}
%
%Similarly, the coalescence parameter \bA~for a nucleus with mass number $A$ and spin $J_{A}$ is generalised by Eq. 6.2 of~\cite{Scheibl:1998tk}.
%For the case of \hethree, the latter becomes Eq. 9 of~\cite{Blum:2017qnn}.
Under the assumption $R_1\approx R_2 \approx R_3 \approx R$ as in \cite{Blum:2017qnn} and by combining Eq. 6.2 in \cite{Scheibl:1998tk} with our Eq.~\ref{eq:CA_general}, 
we obtain:
\begin{equation}\label{eq:master}
	\boxed {  B_{A} = {2J_{A} + 1 \over 2^{A}} {1 \over \sqrt{A}} {1 \over m_{T}^{A-1}} \Bigl({2\pi \over R^2 + ({r_A \over 2})^2 }\Bigr)^{\frac{3}{2}(A-1)} } \;.
\end{equation}
This general formula can be used to compare the predicted $B_{A}$ with data directly.
For small sources, as $R \rightarrow 0$, the coalescence probability is anti-proportional to the harmonic oscillator size parameter, and thus proportional to the depth of the attractive potential in the harmonic oscillator picture (and thus to the nucleus binding energy).
Quite naturally, the allowed momentum difference between the coalescing nucleons is larger for more attractive, i.e. deeper, potentials. 
For a large source where $R \gg r_A$, the coalescence probability is dominated by the classical phase-space separation and, thus, decreases for large distances in configuration space. 

\section{Statistical-thermal approach combined with the blast-wave model}\label{sec:thermal}
In the statistical-thermal approach \cite{Petran:2013dva,Wheaton:2004qb,Andronic:2005yp}, the yields (d$N$/d$y$) of light anti- and hyper-nuclei are very sensitive to $T_{chem}$ due to their large mass and scale approximately as d$N$/d$y \propto (2J_{A} + 1)\exp(-m/T_{chem})$. 
The thermal model implements eigenvolume corrections by fixing the object radius as an external parameter.
We refer to the literature for the extensive discussions on the validity of the eigenvolume correction for light \mbox{(anti-)(hyper-)}nuclei~\cite{Vovchenko:2016mwg} and the relation with the possible production as compact quark bags~\cite{Andronic:2017}.
In contrast to coalescence, the statistical-thermal models provide only \pt-integrated yields. 
Therefore, we use in addition a blast-wave~\cite{Schnedermann:1993ws} parameterization to model the \pt-dependence, with parameters obtained from the simultaneous fit to pion, kaon and proton spectra measured in Pb--Pb collisions by ALICE for several centralities~\cite{Abelev:2013vea}.  
 As discussed in the following section, centrality-dependent parameters allow one to extract spectral distributions for a given source size. 
The object size does not enter in the formulation of the blast-wave model.

The normalisation of the predicted blast-wave spectra for nuclei is fixed using the \pt-integrated deuteron-to-pion ratio and \hethree-to-pion ratio predicted by the GSI-Heidelberg implementation of the statistical-thermal model with $T_{chem} = 156$ MeV, multiplied by the measured pion yield~\cite{Abelev:2013vea}. 
This choice, as opposed to using the ratio to protons, is motivated by the fact that the measured proton yield is seen to be slightly overestimated by the thermal model~\cite{Abelev:2012wca}.
For hyper-triton, the normalisation is extracted from the statistical-thermal model prediction of the strangeness population factor \mbox{$S_{3} = { _{\Lambda}^{3}\mathrm{H} / ^{3}{\mathrm{He}} \over \Lambda/p}$} multiplied by the measured $\Lambda/p$ ratio~\cite{Abelev:2013vea,Abelev:2013xaa} and \hethree~yield~\cite{ALICE:deuteronppPbPb2015}. 
With the resulting spectra, we calculate \bA~for a given \pt/$A$ and compare it with coalescence expectations. 
In the following, we label this model ``thermal+blast-wave''.

\section{Mapping event multiplicity into source size} \label{sec:radiiParamet}
In order to compare the source radius-dependent predictions from coalescence with the centrality-dependent data and with predictions from the thermal+blast-wave model, we map the average charged particle multiplicity density (\avdNdeta) in each centrality (or multiplicity) event class into the system size. 
Experimentally, the source size can be controlled by selecting different collision geometries, i.e. different centralities \cite{Abelev:2013qoq}. 
This mapping is based on the parameterisation
\begin{equation}
R = a \avdNdeta^{1/3} + b\;\;
\label{eq:radiusParameterisation}
\end{equation}
\noindent where $R$ is the source radius, $a = 0.473$ fm and $b$ = 0. 

The value of the empirical parameter $a$ is obtained by tuning the parameterisation such that the measured (anti-)deuteron \btwo~in the most central \PbPb~class falls onto the coalescence prediction. 
In this way, we constrain the coalescence volume with the more differential (anti-)deuteron data and assume that it is the same for all anti- and hyper-nuclei. 
We justify the choice of Eq.~\ref{eq:radiusParameterisation} by identifying the source volume as the effective sub-volume of the whole system that is governed by the homogeneity length of the interacting nucleons (as in \cite{Scheibl:1998tk}) and experimentally accessible with Hanbury-Brown-Twiss (HBT) interferometry~\footnote{To match the Bertsch-Pratt parameterisation ($R_{out}, R_{side}, R_{long}$) employed by experiments, with the Yano-Koonin-Podgoretskii (\rperp, \rpar, $R_{0}$) parameterisation used in the coalescence model,  we identify $\rperp = R_{side}$, $\rpar = R_{long}$ and then take $R = (\rperp^{2}\rpar)^{1/3} \approx (R_{side}^{2}R_{long})^{1/3}$.}.
The HBT radii scale with \avdNdeta$^{1/3}$ and we make the simplifying assumption that this scaling holds across collision systems, which is approximately fulfilled in the data \cite{Adam:2015vna,Adam:2015pya}. 
We also note that the HBT radii, and thus also the source size, depend on the pair average transverse momentum $\langle k_{\mathrm{T}}\rangle$~\cite{Aamodt:2011mr}. 
In contrast to \cite{Blum:2017qnn}, we do not explicitly use the measured HBT radii in our study because using a linear fit to the ALICE HBT data \cite{Abelev:2012sq,Adam:2015vna} would result in a smaller source size ($R \approx$~4~fm) than required by the measured \btwo~to agree with the coalescence prediction ($R \approx$~5.5~fm) in central \PbPb~collisions.
We do however take into account the experimentally observed $\langle k_{\mathrm{T}}\rangle$ dependence of the source size, in contrast to a similar coalescence study reported in \cite{Sun:2018mqq} (for a detailed discussion of possible alternative source volume parameterisations, see the Appendix \ref{appendixA}). 
Production via coalescence could also be investigated by looking at the transverse momentum dependence of $B_{A}$. 
However, the advantage of studying these effects as a function of the multiplicity/centrality is that the system size offers a larger lever arm. 
For a fixed \pt/$A$, \btwo~changes by a factor of about 50 going from pp to central \PbPb~collisions, whereas \btwo~changes by a factor of two going from \pt/$A$~=~0.4~GeV/$c$ to \pt/$A$~=~2.2~GeV/$c$  in most central \PbPb~collisions \cite{ALICE:deuteronppPbPb2015}, and by a factor of less than two in the measured \pt/$A$ range in pp collisions \cite{ALICE:nucleipp2017, Acharya:2019rgc}.

\section{Comparison with data}\label{sec:results}

\begin{figure}[!h]
	\begin{center}
		\includegraphics[width=.95\columnwidth]{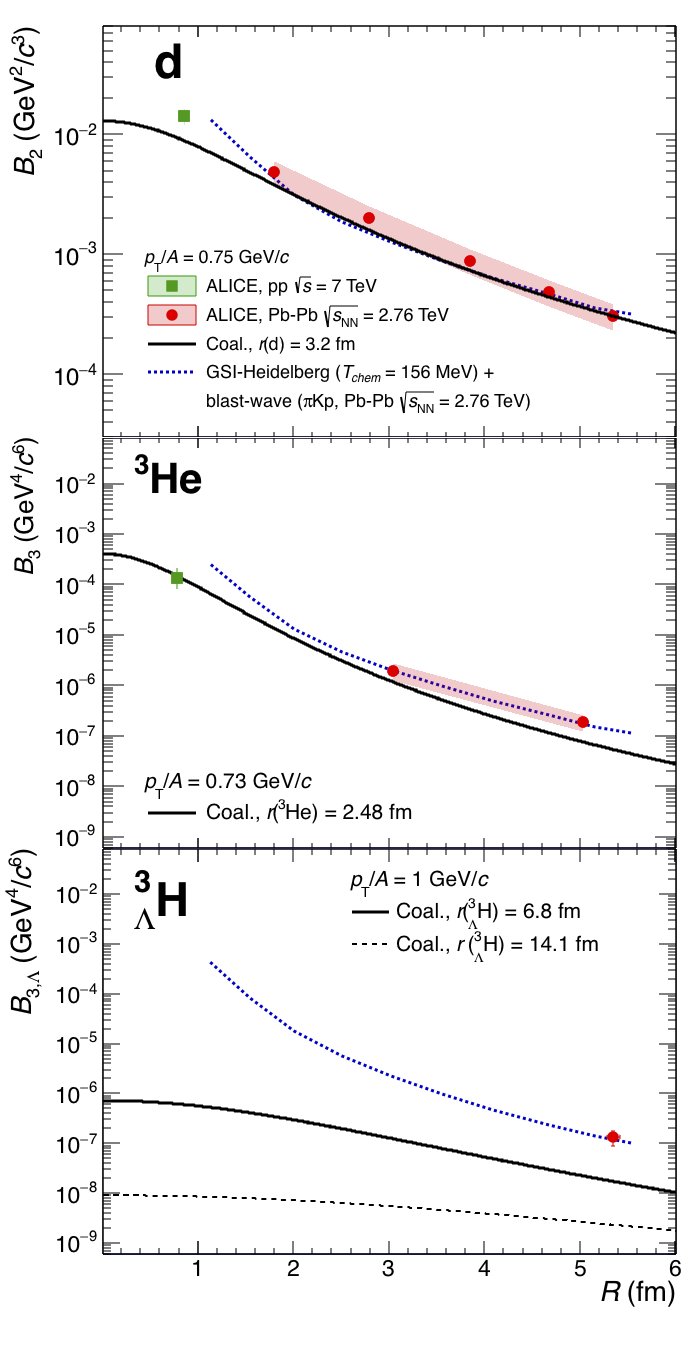}
		\caption{Comparison of the coalescence parameters measured by ALICE (filled symbols) for deuterons (upper panel), \hethree\ (middle panel) and \hthreelambda\ (lower panel) in pp \cite{ALICE:nucleipp2017} and Pb--Pb~\cite{ALICE:deuteronppPbPb2015, Adam:2015yta} collisions with the  thermal+blast-wave model expectations (dotted line) and the coalescence predictions (solid lines). The dashed line in the lower right panel corresponds to the coalescence prediction for the \hthreelambda\ with a larger radius. We have rescaled the inelastic pp collision data in \cite{ALICE:nucleipp2017} to match the so-called INEL$>$0 class by the ratio of \avdNdeta~in these two event classes, see~\cite{Adam:2015gka} for details. 
		}
		\label{fig:CompareThermalAndCoalescence}
	\end{center}
\end{figure}

In Fig.~\ref{fig:CompareThermalAndCoalescence}, the available LHC data for \mbox{(anti-)(hyper-)nuclei} \cite{ALICE:nucleipp2017,ALICE:deuteronppPbPb2015,Adam:2015yta} are compared to coalescence and to the thermal+blast-wave predictions.
For the latter and for data, the radius parameterisation given by Eq.~\ref{eq:radiusParameterisation} is used.
For deuterons, both approaches lead to similar predictions and describe reasonably the data for $R \gtrsim 1.6$ fm. 
For \hethree, the models exhibit a qualitatively similar $R$-dependence but differ by a factor of about 1.5 to 2. 
The currently available data are consistent with both models within 2$\sigma$ to 3$\sigma$, where $\sigma$ is the total uncertainty in the data. 
Both approaches show large differences (a factor of 5 to 6 for central Pb--Pb collisions and a factor of more than 50 for $R < 2$~fm) for the \hthreelambda\ caused by the significantly larger size of \hthreelambda\ with respect to \hethree. 
The only data point available so far in Pb--Pb collisions is in agreement with the thermal+blast-wave model but differs by 6$\sigma$ from our coalescence calculation. This discrepancy might differ for calculations that use a more realistic (non-Gaussian) wave-function.
In \cite{Zhang:2018euf} it is argued that the difference between data and coalescence might be explained by a later formation through coalescence of $\Lambda$s and deuterons. 
A possible difference attributable to the presence of excited states with $J=3/2$ of \hthreelambda~is not considered here, as there is no evidence for its existence~\cite{Mart:1996ay}. 

Most importantly, Fig.~\ref{fig:CompareThermalAndCoalescence} shows that the difference between the two approaches increases with decreasing $R$, highlighting the need for additional multiplicity-differential data to distinguish between the two production scenarios. 
For \hthreelambda~we considered also a prediction from coalescence for a wider wave-function (see Fig. \ref{fig:CompareThermalAndCoalescence}), which results in even lower production probabilities. This behaviour highlights the unique potential to constrain the wave-function of (hyper-)nuclei via precise measurements of the $R$-dependent coalescence parameter. 
The curves presented here explicitly allow for an estimate of the hyper-triton production in pp collisions, which is expected to be suppressed by about two orders of magnitude with respect to that of \hethree, making this measurement a prime candidate for future experimental studies.

\section{(Anti-)triton production}
For isobars with the same spin but different wave-functions, like \tritium~(\antitritium)~and \hethree~(\antihethree), Eq.~\ref{eq:master} provides a relation between the relative coalescence probabilities as the corollary:

\begin{equation} \label{eq:rho}
\rho (\pt) \equiv \frac{B_{A}(\tritium)}{B_{A}(\hethree)}(\pt) = \bigg( \frac{R(\pt)^{2} + \frac{r_{\hethree}^{2}}{4}} {R(\pt)^{2} + \frac{r_{\tritium}^{2}}{4}} \bigg)^{3} \;\; .
\end{equation}

\begin{figure}[!h]
	\begin{center}
		\includegraphics[width=\columnwidth]{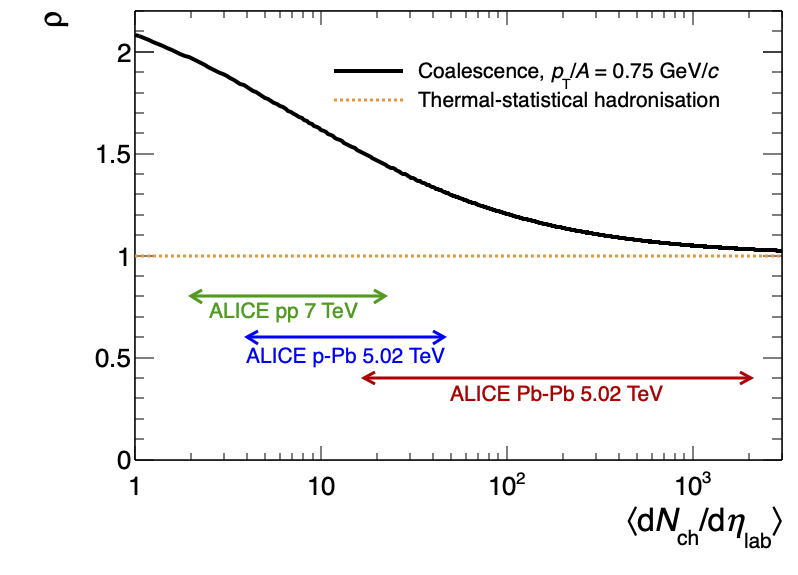}
		\caption{Yield of \tritium~relative to \hethree~for \pt/$A$~=~0.75~\GeVc~as defined by Eq.~\ref{eq:rho} in the coalescence picture (black line). The source radius has been mapped into the average charged particle multiplicity by using the parameterisation constrained to the ALICE \btwo~measurement described in Sec.~\ref{sec:radiiParamet}. The dashed orange line represents the expectation from statistical hadronisation models. Colored arrows highlight the multiplicity range spanned by ALICE measurements in different collision systems.}
		\label{fig:rhoratio}
	\end{center}
\end{figure}

\noindent Under the assumption that the distributions of protons and neutrons are identical, this is equivalent to the ratio of \tritium~yield relative to that of \hethree~(or \antitritium/\antihethree) at a given transverse momentum. It is to be noted that the transverse momentum dependence of this ratio originates solely from the \pt-dependence of the source volume. 
The ratio $\rho$ is reported in Fig.~\ref{fig:rhoratio} for \pt/$A$~=~0.75 \GeVc~of the two isobars and as a function of the average charged particle multiplicity. 
The multiplicity is obtained from the system radius by inverting Eq.~\ref{eq:radiusParameterisation}.
The coalescence model predicts a dependence of $\rho$ on the system size and thus on the charged particle multiplicity. In particular, the production of \tritium~(\antitritium) is predicted by our model to be more abundant by up to a factor of two than the production of \hethree~(\antihethree) in small systems. Such an excess of (anti-)tritons is not expected in thermal-statistical hadronisation. Since the two isobars have the same mass, spin, and baryon number, the Grand Canonical thermal model~\cite{Andronic:2010qu,Andronic:2017} predicts a \tritium/\hethree~ratio equal to unity. The same holds true when taking into account the potential suppression of \mbox{(anti-)nucleus} yields due to explicit conservation of the baryon number, as in Canonical Statistical Model calculations~\cite{Vovchenko:2018fiy}.
Existing (anti-)triton measurements in pp collisions~\cite{ALICE:nucleipp2017} are limited by statistical precision and thus do not yet allow for a distinction between the two scenarios.

\section{Conclusions and outlook}
We summarise our main conclusions as follows:
\begin{enumerate}
	\item For the production of $A=2$ and $A=3$ \mbox{(anti-)nuclei} in heavy-ion collisions, thermal+blast-wave and coalescence models give similar predictions for a source volume that is constrained by experimental data on d, $\bar{\mathrm{d}}$~production in central Pb--Pb collisions.
	\item For hyper-triton, the two models give very different predictions as a function of source volume. In particular, the yield of hyper-triton appears to be suppressed by about two orders of magnitude in pp collisions with respect to \hethree~due to its wider wave-function. 
	\item In \PbPb~collisions, the very limited number of data available favours the thermal+blast-wave model prediction within our assumptions.
	\item Systematic measurements in pp, p--Pb, and Pb--Pb collisions at LHC energies have a unique potential to clarify the production mechanism and the nature of composite QCD objects. Ideally, such measurements are accompanied by systematic measurements of the HBT radii in the same multiplicity/centrality classes and collision systems.
	\item Our findings suggest a clear experimental path to be pursued with high-precision measurements at the upcoming phase of the LHC in the next ten years, which will finally provide sufficient integrated luminosity for the studies proposed here~\cite{Citron:2018lsq}.
% will provide a unique opportunity for the final understanding of \mbox{(anti-)(hyper-)nuclei} production.

\end{enumerate}
As our study is deliberately based on simplified assumptions that allow for a completely analytical treatment of the problem, future studies should be based on more realistic approximations (in particular the wave-function), which require numerical calculations. 
We plan to extend our study to explore further the \pt~dependence as well as to investigate $A=4$ systems and more exotic QCD objects like the X(3872)~\cite{Esposito:2015fsa, Cho:2017dcy}. 
If the X(3872) corresponds to a loosely bound $\overline{D}^{*0}$--$D^{0}$ molecule, the rms of its wave-function can be as large as $4.9^{+13.4}_{-1.4}$~fm \cite{Artoisenet:2010uu}. Thus, its possible production via coalescence in pp collisions would be subject to a similar suppression as the hyper-triton. 
Setting a final word on the production mechanisms also has a broader application in astrophysics and dark-matter searches, by representing an essential input for the measurement of (anti-)nuclei in space within ongoing \cite{Alcaraz:2000ss, Poulin:2018wzu} and future \cite{AMS100, Aramaki:2015laa} experiments.

\appendix
\section{Comparison of alternative source volume parameterisations} \label{appendixA} 
\begin{figure}[!h]
	\begin{center}
		\includegraphics[width=\columnwidth]{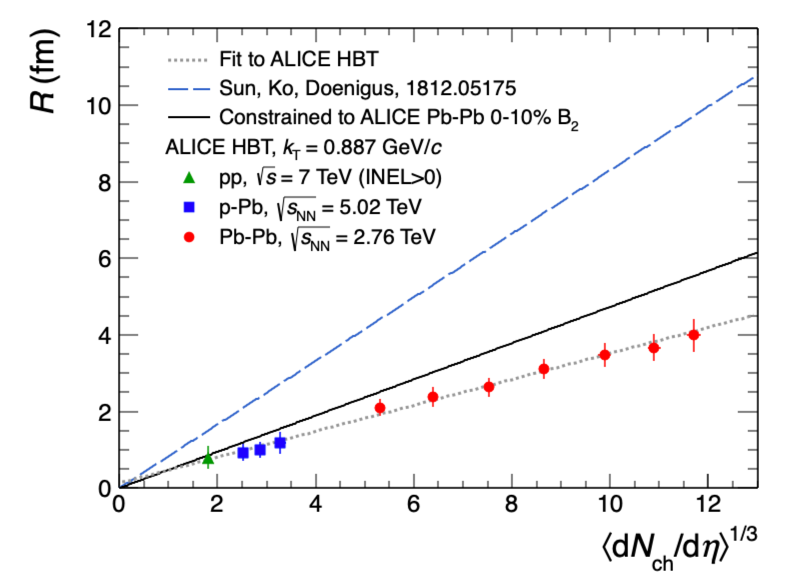}
		\caption{Comparison of different source volume parameterisations. The dotted gray line is the result of a linear fit to the ALICE HBT data~\cite{Adam:2015vna, Adam:2015pya, Abelev:2012sq}, according to Eq.~\ref{eq:radiusParameterisation} with $a$~=~0.339~fm and $b$~=~0.128~fm.  
		The dashed blue line represents the parameterisation used in \cite{Sun:2018mqq}, corresponding to $a$~=~0.83~fm and $b$~=~0.
		The solid black line represents the parameterisation used in our study ($a$~=~0.473~fm and $b$~=~0) and obtained by a constraint from the \btwo~measured in central (0-10$\%$) \PbPb~collisions.}
		\label{fig:sourceVolume}
	\end{center}
\end{figure}

For the considerations discussed in this work and in similar reports, one crucial aspect has to be taken into account, namely how the source radius is parameterised as a function of the multiplicity and transverse momentum.
As discussed in Sec.~\ref{sec:radiiParamet}, for our main result we rely on the parameterisation given by Eq.~\ref{eq:radiusParameterisation}, which assumes a linear dependence between the cubic root of the measured average charged particle multiplicity density and the radius of the source. 
In order to extract the $a$ and $b$ coefficients of Eq.~\ref{eq:radiusParameterisation}, several solutions can be adopted. 
One possibility is to fix the coefficients by fitting the available HBT data from ALICE~\cite{Adam:2015vna, Adam:2015pya, Abelev:2012sq} with Eq.~\ref{eq:radiusParameterisation}, as shown in Fig.~\ref{fig:sourceVolume} by the data points fitted with the dotted gray line. 
In order to preserve the momentum dependence of the source volume, the highest available $k_{\mathrm{T}}$ ($\approx$ 0.9 \GeVc) from pion HBT is chosen as it is closest in $m_{\mathrm{T}}$ to the lowest transverse momentum per nucleon ($p_{\mathrm{T}} \approx$ 0.8 \GeVc) accessible by ALICE for the measurement of nucleus production. 
Ideally, one would use the proton femtoscopic radii, but given the availability of these measurements in only some of the collision systems and centralities, we assume that $m_{\mathrm{T}}$-scaling holds.
The fit to the ALICE data results in $a$~=~0.339~fm and $b$~=~0.128~fm.
However, this approach does not describe the deuteron \btwo~data: we observe that in order for the measured \btwo~to agree with the coalescence prediction in most central \PbPb~collisions, a larger source radius of $R \approx 5.5$ fm is required instead of the $R \approx 4$ fm resulting from the HBT fit.
This might also indicate that the volume relevant for the coalescence process is slightly larger than the homogeneity volume that can be extracted from HBT studies.
Because of this, we chose to constrain the parameterisation to the measured \btwo~in central \PbPb~collisions for \pt~=~0.75 \GeVc~as well as to the origin ($b = 0$). 
The resulting parameterisation, with $a = 0.473$ fm, is represented in Fig.~\ref{fig:sourceVolume} with a solid black line, and it is used for the results of our study discussed in Sec.~\ref{sec:results}.

The authors of \cite{Sun:2018mqq} assume the same functional dependence as we do in Eq.~\ref{eq:radiusParameterisation}. 
However, their approach deviates from ours as they neglect the momentum dependence of the source volume and constrain it with HBT data at low momentum ($k_{\rm T} = 0.25$ \GeVc). 
The parameterisation from \cite{Sun:2018mqq} ($a$~=~0.83~fm and $b$~=~0) is shown for comparison in Fig.~\ref{fig:sourceVolume} as the dashed blue line.
While this choice allows for a good description of \pt-integrated yield ratios (e.g., d/p, \hethree/p, \hthreelambda/$\Lambda$)~\cite{Sun:2018mqq}, using this parameterisation for the \pt-differential study would lead to a significantly lower coalescence probability than measured, caused by the much larger source volume (e.g., a \btwo~lower by a factor of 5.6 for \pt~=~0.75~\GeVc, corresponding to a factor of two larger radius). 
In other words, by integrating over the transverse momentum, one loses sensitivity to the \pt-dependence of the coalescence probability as the object radius becomes dominated by the large source volumina for small $k_{\rm T}$ (see Eq.~\ref{eq:master}).

\bigskip
 
\begin{acknowledgments}
We would like to thank Kfir Blum for inspiring this work. We thank U. Heinz for the useful discussions and the clarification on the equivalence of the Bertsch-Pratt and Yano-Koonin-Podgoretskii parameterisations of the HBT radii. We further acknowledge discussions with Benjamin Doenigus, in particular about the production of \hthreelambda\ in pp collisions, and with Eulogio Serradilla Rodriguez. In addition, we would like to thank Juergen Schukraft, Peter Braun-Munzinger, Marco Van Leuween, Maximiliano Puccio, Roman Lietava, Natasha Sharma, Sebastian Hornung and the colleagues from the ALICE Collaboration for their valuable input. 
\end{acknowledgments}

\bibliography{main_aps}

%merlin.mbs apsrev4-1.bst 2010-07-25 4.21a (PWD, AO, DPC) hacked
%Control: key (0)
%Control: author (8) initials jnrlst
%Control: editor formatted (1) identically to author
%Control: production of article title (-1) disabled
%Control: page (0) single
%Control: year (1) truncated
%Control: production of eprint (0) enabled
\begin{thebibliography}{61}%
\makeatletter
\providecommand \@ifxundefined [1]{%
 \@ifx{#1\undefined}
}%
\providecommand \@ifnum [1]{%
 \ifnum #1\expandafter \@firstoftwo
 \else \expandafter \@secondoftwo
 \fi
}%
\providecommand \@ifx [1]{%
 \ifx #1\expandafter \@firstoftwo
 \else \expandafter \@secondoftwo
 \fi
}%
\providecommand \natexlab [1]{#1}%
\providecommand \enquote  [1]{``#1''}%
\providecommand \bibnamefont  [1]{#1}%
\providecommand \bibfnamefont [1]{#1}%
\providecommand \citenamefont [1]{#1}%
\providecommand \href@noop [0]{\@secondoftwo}%
\providecommand \href [0]{\begingroup \@sanitize@url \@href}%
\providecommand \@href[1]{\@@startlink{#1}\@@href}%
\providecommand \@@href[1]{\endgroup#1\@@endlink}%
\providecommand \@sanitize@url [0]{\catcode `\\12\catcode `\$12\catcode
  `\&12\catcode `\#12\catcode `\^12\catcode `\_12\catcode `\%12\relax}%
\providecommand \@@startlink[1]{}%
\providecommand \@@endlink[0]{}%
\providecommand \url  [0]{\begingroup\@sanitize@url \@url }%
\providecommand \@url [1]{\endgroup\@href {#1}{\urlprefix }}%
\providecommand \urlprefix  [0]{URL }%
\providecommand \Eprint [0]{\href }%
\providecommand \doibase [0]{http://dx.doi.org/}%
\providecommand \selectlanguage [0]{\@gobble}%
\providecommand \bibinfo  [0]{\@secondoftwo}%
\providecommand \bibfield  [0]{\@secondoftwo}%
\providecommand \translation [1]{[#1]}%
\providecommand \BibitemOpen [0]{}%
\providecommand \bibitemStop [0]{}%
\providecommand \bibitemNoStop [0]{.\EOS\space}%
\providecommand \EOS [0]{\spacefactor3000\relax}%
\providecommand \BibitemShut  [1]{\csname bibitem#1\endcsname}%
\let\auto@bib@innerbib\@empty
%</preamble>
\bibitem [{\citenamefont {Adam}\ \emph
  {et~al.}(2016{\natexlab{a}})\citenamefont {Adam} \emph
  {et~al.}}]{Adam:2015vna}%
  \BibitemOpen
  \bibfield  {author} {\bibinfo {author} {\bibfnamefont {J.}~\bibnamefont
  {Adam}} \emph {et~al.} (\bibinfo {collaboration} {ALICE}),\ }\href {\doibase
  10.1103/PhysRevC.93.024905} {\bibfield  {journal} {\bibinfo  {journal} {Phys.
  Rev.}\ }\textbf {\bibinfo {volume} {C93}},\ \bibinfo {pages} {024905}
  (\bibinfo {year} {2016}{\natexlab{a}})},\ \Eprint
  {http://arxiv.org/abs/1507.06842} {arXiv:1507.06842 [nucl-ex]} \BibitemShut
  {NoStop}%
%%CITATION = ARXIV:1507.06842;%%
\bibitem [{\citenamefont {Nemura}\ \emph {et~al.}(2000)\citenamefont {Nemura},
  \citenamefont {Suzuki}, \citenamefont {Fujiwara},\ and\ \citenamefont
  {Nakamoto}}]{Nemura:1999qp}%
  \BibitemOpen
  \bibfield  {author} {\bibinfo {author} {\bibfnamefont {H.}~\bibnamefont
  {Nemura}}, \bibinfo {author} {\bibfnamefont {Y.}~\bibnamefont {Suzuki}},
  \bibinfo {author} {\bibfnamefont {Y.}~\bibnamefont {Fujiwara}}, \ and\
  \bibinfo {author} {\bibfnamefont {C.}~\bibnamefont {Nakamoto}},\ }\href
  {\doibase 10.1143/PTP.103.929} {\bibfield  {journal} {\bibinfo  {journal}
  {Prog. Theor. Phys.}\ }\textbf {\bibinfo {volume} {103}},\ \bibinfo {pages}
  {929} (\bibinfo {year} {2000})},\ \Eprint
  {http://arxiv.org/abs/nucl-th/9912065} {arXiv:nucl-th/9912065 [nucl-th]}
  \BibitemShut {NoStop}%
%%CITATION = NUCL-TH/9912065;%%
\bibitem [{\citenamefont {Butler}\ and\ \citenamefont
  {Pearson}(1963)}]{Butler:1963}%
  \BibitemOpen
  \bibfield  {author} {\bibinfo {author} {\bibfnamefont {S.~T.}\ \bibnamefont
  {Butler}}\ and\ \bibinfo {author} {\bibfnamefont {C.~A.}\ \bibnamefont
  {Pearson}},\ }\href {\doibase 10.1103/PhysRev.129.836} {\bibfield  {journal}
  {\bibinfo  {journal} {Phys. Rev.}\ }\textbf {\bibinfo {volume} {129}},\
  \bibinfo {pages} {836} (\bibinfo {year} {1963})}\BibitemShut {NoStop}%
%%CITATION = PHRVA,129,836;%%
\bibitem [{\citenamefont {Kapusta}(1980)}]{Kapusta:1980}%
  \BibitemOpen
  \bibfield  {author} {\bibinfo {author} {\bibfnamefont {J.~I.}\ \bibnamefont
  {Kapusta}},\ }\href {\doibase 10.1103/PhysRevC.21.1301} {\bibfield  {journal}
  {\bibinfo  {journal} {Phys. Rev.}\ }\textbf {\bibinfo {volume} {C21}},\
  \bibinfo {pages} {1301} (\bibinfo {year} {1980})}\BibitemShut {NoStop}%
%%CITATION = PHRVA,C21,1301;%%
\bibitem [{\citenamefont {Sato}\ and\ \citenamefont
  {Yazaki}(1981)}]{Sato:1981ez}%
  \BibitemOpen
  \bibfield  {author} {\bibinfo {author} {\bibfnamefont {H.}~\bibnamefont
  {Sato}}\ and\ \bibinfo {author} {\bibfnamefont {K.}~\bibnamefont {Yazaki}},\
  }\href {\doibase 10.1016/0370-2693(81)90976-X} {\bibfield  {journal}
  {\bibinfo  {journal} {Phys. Lett.}\ }\textbf {\bibinfo {volume} {B98}},\
  \bibinfo {pages} {153} (\bibinfo {year} {1981})}\BibitemShut {NoStop}%
%%CITATION = PHLTA,98B,153;%%
\bibitem [{\citenamefont {Nagle}\ \emph {et~al.}(1996)\citenamefont {Nagle},
  \citenamefont {Kumar}, \citenamefont {Kusnezov}, \citenamefont {Sorge},\ and\
  \citenamefont {Mattiello}}]{Nagle:1996vp}%
  \BibitemOpen
  \bibfield  {author} {\bibinfo {author} {\bibfnamefont {J.~L.}\ \bibnamefont
  {Nagle}}, \bibinfo {author} {\bibfnamefont {B.~S.}\ \bibnamefont {Kumar}},
  \bibinfo {author} {\bibfnamefont {D.}~\bibnamefont {Kusnezov}}, \bibinfo
  {author} {\bibfnamefont {H.}~\bibnamefont {Sorge}}, \ and\ \bibinfo {author}
  {\bibfnamefont {R.}~\bibnamefont {Mattiello}},\ }\href {\doibase
  10.1103/PhysRevC.53.367} {\bibfield  {journal} {\bibinfo  {journal} {Phys.
  Rev.}\ }\textbf {\bibinfo {volume} {C53}},\ \bibinfo {pages} {367} (\bibinfo
  {year} {1996})}\BibitemShut {NoStop}%
%%CITATION = PHRVA,C53,367;%%
\bibitem [{\citenamefont {Scheibl}\ and\ \citenamefont
  {Heinz}(1999)}]{Scheibl:1998tk}%
  \BibitemOpen
  \bibfield  {author} {\bibinfo {author} {\bibfnamefont {R.}~\bibnamefont
  {Scheibl}}\ and\ \bibinfo {author} {\bibfnamefont {U.~W.}\ \bibnamefont
  {Heinz}},\ }\href {\doibase 10.1103/PhysRevC.59.1585} {\bibfield  {journal}
  {\bibinfo  {journal} {Phys. Rev.}\ }\textbf {\bibinfo {volume} {C59}},\
  \bibinfo {pages} {1585} (\bibinfo {year} {1999})},\ \Eprint
  {http://arxiv.org/abs/nucl-th/9809092} {arXiv:nucl-th/9809092 [nucl-th]}
  \BibitemShut {NoStop}%
%%CITATION = NUCL-TH/9809092;%%
\bibitem [{\citenamefont {Cho}\ \emph {et~al.}(2017)\citenamefont {Cho} \emph
  {et~al.}}]{Cho:2017dcy}%
  \BibitemOpen
  \bibfield  {author} {\bibinfo {author} {\bibfnamefont {S.}~\bibnamefont
  {Cho}} \emph {et~al.} (\bibinfo {collaboration} {ExHIC}),\ }\href {\doibase
  10.1016/j.ppnp.2017.02.002} {\bibfield  {journal} {\bibinfo  {journal} {Prog.
  Part. Nucl. Phys.}\ }\textbf {\bibinfo {volume} {95}},\ \bibinfo {pages}
  {279} (\bibinfo {year} {2017})},\ \Eprint {http://arxiv.org/abs/1702.00486}
  {arXiv:1702.00486 [nucl-th]} \BibitemShut {NoStop}%
%%CITATION = ARXIV:1702.00486;%%
\bibitem [{\citenamefont {Blum}\ \emph {et~al.}(2017)\citenamefont {Blum},
  \citenamefont {Ng}, \citenamefont {Sato},\ and\ \citenamefont
  {Takimoto}}]{Blum:2017qnn}%
  \BibitemOpen
  \bibfield  {author} {\bibinfo {author} {\bibfnamefont {K.}~\bibnamefont
  {Blum}}, \bibinfo {author} {\bibfnamefont {K.~C.~Y.}\ \bibnamefont {Ng}},
  \bibinfo {author} {\bibfnamefont {R.}~\bibnamefont {Sato}}, \ and\ \bibinfo
  {author} {\bibfnamefont {M.}~\bibnamefont {Takimoto}},\ }\href {\doibase
  10.1103/PhysRevD.96.103021} {\bibfield  {journal} {\bibinfo  {journal} {Phys.
  Rev.}\ }\textbf {\bibinfo {volume} {D96}},\ \bibinfo {pages} {103021}
  (\bibinfo {year} {2017})},\ \Eprint {http://arxiv.org/abs/1704.05431}
  {arXiv:1704.05431 [astro-ph.HE]} \BibitemShut {NoStop}%
%%CITATION = ARXIV:1704.05431;%%
\bibitem [{\citenamefont {Bazak}\ and\ \citenamefont
  {Mrowczynski}(2018)}]{Bazak:2018hgl}%
  \BibitemOpen
  \bibfield  {author} {\bibinfo {author} {\bibfnamefont {S.}~\bibnamefont
  {Bazak}}\ and\ \bibinfo {author} {\bibfnamefont {S.}~\bibnamefont
  {Mrowczynski}},\ }\href {\doibase 10.1142/S0217732318501420} {\bibfield
  {journal} {\bibinfo  {journal} {Mod. Phys. Lett.}\ }\textbf {\bibinfo
  {volume} {A33}},\ \bibinfo {pages} {1850142} (\bibinfo {year} {2018})},\
  \Eprint {http://arxiv.org/abs/1802.08212} {arXiv:1802.08212 [nucl-th]}
  \BibitemShut {NoStop}%
%%CITATION = ARXIV:1802.08212;%%
\bibitem [{\citenamefont {Zhao}\ \emph {et~al.}(2018)\citenamefont {Zhao},
  \citenamefont {Zhu}, \citenamefont {Zheng}, \citenamefont {Ko},\ and\
  \citenamefont {Song}}]{Zhao:2018lyf}%
  \BibitemOpen
  \bibfield  {author} {\bibinfo {author} {\bibfnamefont {W.}~\bibnamefont
  {Zhao}}, \bibinfo {author} {\bibfnamefont {L.}~\bibnamefont {Zhu}}, \bibinfo
  {author} {\bibfnamefont {H.}~\bibnamefont {Zheng}}, \bibinfo {author}
  {\bibfnamefont {C.~M.}\ \bibnamefont {Ko}}, \ and\ \bibinfo {author}
  {\bibfnamefont {H.}~\bibnamefont {Song}},\ }\href {\doibase
  10.1103/PhysRevC.98.054905} {\bibfield  {journal} {\bibinfo  {journal} {Phys.
  Rev.}\ }\textbf {\bibinfo {volume} {C98}},\ \bibinfo {pages} {054905}
  (\bibinfo {year} {2018})},\ \Eprint {http://arxiv.org/abs/1807.02813}
  {arXiv:1807.02813 [nucl-th]} \BibitemShut {NoStop}%
%%CITATION = ARXIV:1807.02813;%%
\bibitem [{\citenamefont {Andronic}\ \emph {et~al.}(2011)\citenamefont
  {Andronic}, \citenamefont {Braun-Munzinger}, \citenamefont {Stachel},\ and\
  \citenamefont {Stocker}}]{Andronic:2010qu}%
  \BibitemOpen
  \bibfield  {author} {\bibinfo {author} {\bibfnamefont {A.}~\bibnamefont
  {Andronic}}, \bibinfo {author} {\bibfnamefont {P.}~\bibnamefont
  {Braun-Munzinger}}, \bibinfo {author} {\bibfnamefont {J.}~\bibnamefont
  {Stachel}}, \ and\ \bibinfo {author} {\bibfnamefont {H.}~\bibnamefont
  {Stocker}},\ }\href {\doibase https://doi.org/10.1016/j.physletb.2011.01.053}
  {\bibfield  {journal} {\bibinfo  {journal} {Physics Letters B}\ }\textbf
  {\bibinfo {volume} {697}},\ \bibinfo {pages} {203 } (\bibinfo {year}
  {2011})}\BibitemShut {NoStop}%
\bibitem [{\citenamefont {Andronic}\ \emph {et~al.}(2018)\citenamefont
  {Andronic}, \citenamefont {Braun-Munzinger}, \citenamefont {Redlich},\ and\
  \citenamefont {Stachel}}]{Andronic:2017}%
  \BibitemOpen
  \bibfield  {author} {\bibinfo {author} {\bibfnamefont {A.}~\bibnamefont
  {Andronic}}, \bibinfo {author} {\bibfnamefont {P.}~\bibnamefont
  {Braun-Munzinger}}, \bibinfo {author} {\bibfnamefont {K.}~\bibnamefont
  {Redlich}}, \ and\ \bibinfo {author} {\bibfnamefont {J.}~\bibnamefont
  {Stachel}},\ }\href@noop {} {\bibfield  {journal} {\bibinfo  {journal}
  {Nature}\ }\textbf {\bibinfo {volume} {561}},\ \bibinfo {pages} {321}
  (\bibinfo {year} {2018})},\ \Eprint {http://arxiv.org/abs/1710.09425}
  {arXiv:1710.09425 [nucl-th]} \BibitemShut {NoStop}%
\bibitem [{\citenamefont {Abelev}\ \emph
  {et~al.}(2013{\natexlab{a}})\citenamefont {Abelev} \emph
  {et~al.}}]{Abelev:2013vea}%
  \BibitemOpen
  \bibfield  {author} {\bibinfo {author} {\bibfnamefont {B.}~\bibnamefont
  {Abelev}} \emph {et~al.} (\bibinfo {collaboration} {ALICE}),\ }\href
  {\doibase 10.1103/PhysRevC.88.044910} {\bibfield  {journal} {\bibinfo
  {journal} {Phys. Rev.}\ }\textbf {\bibinfo {volume} {C88}},\ \bibinfo {pages}
  {044910} (\bibinfo {year} {2013}{\natexlab{a}})},\ \Eprint
  {http://arxiv.org/abs/1303.0737} {arXiv:1303.0737 [hep-ex]} \BibitemShut
  {NoStop}%
%%CITATION = ARXIV:1303.0737;%%
\bibitem [{\citenamefont {Garcilazo}(1982)}]{Garcilazo:1982yc}%
  \BibitemOpen
  \bibfield  {author} {\bibinfo {author} {\bibfnamefont {H.}~\bibnamefont
  {Garcilazo}},\ }\href {\doibase 10.1103/PhysRevLett.48.577} {\bibfield
  {journal} {\bibinfo  {journal} {Phys. Rev. Lett.}\ }\textbf {\bibinfo
  {volume} {48}},\ \bibinfo {pages} {577} (\bibinfo {year} {1982})}\BibitemShut
  {NoStop}%
%%CITATION = PRLTA,48,577;%%
\bibitem [{\citenamefont {Bass}\ \emph {et~al.}(1998)\citenamefont {Bass} \emph
  {et~al.}}]{Bass:1998ca}%
  \BibitemOpen
  \bibfield  {author} {\bibinfo {author} {\bibfnamefont {S.~A.}\ \bibnamefont
  {Bass}} \emph {et~al.},\ }\href {\doibase 10.1016/S0146-6410(98)00058-1}
  {\bibfield  {journal} {\bibinfo  {journal} {Prog. Part. Nucl. Phys.}\
  }\textbf {\bibinfo {volume} {41}},\ \bibinfo {pages} {255} (\bibinfo {year}
  {1998})},\ \Eprint {http://arxiv.org/abs/nucl-th/9803035}
  {arXiv:nucl-th/9803035 [nucl-th]} \BibitemShut {NoStop}%
%%CITATION = NUCL-TH/9803035;%%
\bibitem [{\citenamefont {Schukraft}(2017)}]{Schukraft:2017nbn}%
  \BibitemOpen
  \bibfield  {author} {\bibinfo {author} {\bibfnamefont {J.}~\bibnamefont
  {Schukraft}},\ }\bibfield  {booktitle} {\emph {\bibinfo {booktitle}
  {{Proceedings, 26th International Conference on Ultra-relativistic
  Nucleus-Nucleus Collisions (Quark Matter 2017): Chicago, Illinois, USA,
  February 5-11, 2017}}},\ }\href {\doibase 10.1016/j.nuclphysa.2017.05.036}
  {\bibfield  {journal} {\bibinfo  {journal} {Nucl. Phys.}\ }\textbf {\bibinfo
  {volume} {A967}},\ \bibinfo {pages} {1} (\bibinfo {year} {2017})},\ \Eprint
  {http://arxiv.org/abs/1705.02646} {arXiv:1705.02646 [hep-ex]} \BibitemShut
  {NoStop}%
%%CITATION = ARXIV:1705.02646;%%
\bibitem [{\citenamefont {Oliinychenko}\ \emph {et~al.}(2018)\citenamefont
  {Oliinychenko}, \citenamefont {Pang}, \citenamefont {Elfner},\ and\
  \citenamefont {Koch}}]{Oliinychenko:2018ugs}%
  \BibitemOpen
  \bibfield  {author} {\bibinfo {author} {\bibfnamefont {D.}~\bibnamefont
  {Oliinychenko}}, \bibinfo {author} {\bibfnamefont {L.-G.}\ \bibnamefont
  {Pang}}, \bibinfo {author} {\bibfnamefont {H.}~\bibnamefont {Elfner}}, \ and\
  \bibinfo {author} {\bibfnamefont {V.}~\bibnamefont {Koch}},\ }\href@noop {}
  {\  (\bibinfo {year} {2018})},\ \Eprint {http://arxiv.org/abs/1809.03071}
  {arXiv:1809.03071 [hep-ph]} \BibitemShut {NoStop}%
%%CITATION = ARXIV:1809.03071;%%
\bibitem [{\citenamefont {Acharya}\ \emph {et~al.}(2017)\citenamefont {Acharya}
  \emph {et~al.}}]{Acharya:2017dmc}%
  \BibitemOpen
  \bibfield  {author} {\bibinfo {author} {\bibfnamefont {S.}~\bibnamefont
  {Acharya}} \emph {et~al.} (\bibinfo {collaboration} {ALICE}),\ }\href
  {\doibase 10.1140/epjc/s10052-017-5222-x} {\bibfield  {journal} {\bibinfo
  {journal} {Eur. Phys. J.}\ }\textbf {\bibinfo {volume} {C77}},\ \bibinfo
  {pages} {658} (\bibinfo {year} {2017})},\ \Eprint
  {http://arxiv.org/abs/1707.07304} {arXiv:1707.07304 [nucl-ex]} \BibitemShut
  {NoStop}%
%%CITATION = ARXIV:1707.07304;%%
\bibitem [{\citenamefont {Puccio}(2019)}]{Puccio:2019oyd}%
  \BibitemOpen
  \bibfield  {author} {\bibinfo {author} {\bibfnamefont {M.}~\bibnamefont
  {Puccio}} (\bibinfo {collaboration} {ALICE}),\ }\bibfield  {booktitle} {\emph
  {\bibinfo {booktitle} {{Proceedings, 27th International Conference on
  Ultrarelativistic Nucleus-Nucleus Collisions (Quark Matter 2018): Venice,
  Italy, May 14-19, 2018}}},\ }\href {\doibase 10.1016/j.nuclphysa.2018.10.043}
  {\bibfield  {journal} {\bibinfo  {journal} {Nucl. Phys.}\ }\textbf {\bibinfo
  {volume} {A982}},\ \bibinfo {pages} {447} (\bibinfo {year}
  {2019})}\BibitemShut {NoStop}%
%%CITATION = NUPHA,A982,447;%%
\bibitem [{\citenamefont {Castorina}\ and\ \citenamefont
  {Satz}(2019)}]{Castorina:2019pnb}%
  \BibitemOpen
  \bibfield  {author} {\bibinfo {author} {\bibfnamefont {P.}~\bibnamefont
  {Castorina}}\ and\ \bibinfo {author} {\bibfnamefont {H.}~\bibnamefont
  {Satz}},\ }\href@noop {} {\  (\bibinfo {year} {2019})},\ \Eprint
  {http://arxiv.org/abs/1901.10407} {arXiv:1901.10407 [hep-ph]} \BibitemShut
  {NoStop}%
%%CITATION = ARXIV:1901.10407;%%
\bibitem [{\citenamefont {Xu}\ and\ \citenamefont {Rapp}(2018)}]{Xu:2018jff}%
  \BibitemOpen
  \bibfield  {author} {\bibinfo {author} {\bibfnamefont {X.}~\bibnamefont
  {Xu}}\ and\ \bibinfo {author} {\bibfnamefont {R.}~\bibnamefont {Rapp}},\
  }\href@noop {} {\  (\bibinfo {year} {2018})},\ \Eprint
  {http://arxiv.org/abs/1809.04024} {arXiv:1809.04024 [nucl-th]} \BibitemShut
  {NoStop}%
%%CITATION = ARXIV:1809.04024;%%
\bibitem [{\citenamefont {Heinz}()}]{HeinzTorino}%
  \BibitemOpen
  \bibfield  {author} {\bibinfo {author} {\bibfnamefont {U.}~\bibnamefont
  {Heinz}},\ }\href@noop {} {\enquote {\bibinfo {title} {{Coalescence model
  involving HBT and flow}},}\ }\bibinfo {howpublished} {{Presentation at EMMI
  Workshop in Torino, November 2017}}\BibitemShut {NoStop}%
\bibitem [{\citenamefont {Abelev}\ \emph {et~al.}(2015)\citenamefont {Abelev}
  \emph {et~al.}}]{Abelev:2014uua}%
  \BibitemOpen
  \bibfield  {author} {\bibinfo {author} {\bibfnamefont {B.~B.}\ \bibnamefont
  {Abelev}} \emph {et~al.} (\bibinfo {collaboration} {ALICE}),\ }\href
  {\doibase 10.1103/PhysRevC.91.024609} {\bibfield  {journal} {\bibinfo
  {journal} {Phys. Rev.}\ }\textbf {\bibinfo {volume} {C91}},\ \bibinfo {pages}
  {024609} (\bibinfo {year} {2015})},\ \Eprint {http://arxiv.org/abs/1404.0495}
  {arXiv:1404.0495 [nucl-ex]} \BibitemShut {NoStop}%
%%CITATION = ARXIV:1404.0495;%%
\bibitem [{\citenamefont {Van Der~Leun}\ and\ \citenamefont
  {Alderliesten}(1982)}]{VanDerLeun:1982bhg}%
  \BibitemOpen
  \bibfield  {author} {\bibinfo {author} {\bibfnamefont {C.}~\bibnamefont {Van
  Der~Leun}}\ and\ \bibinfo {author} {\bibfnamefont {C.}~\bibnamefont
  {Alderliesten}},\ }\href {\doibase 10.1016/0375-9474(82)90105-1} {\bibfield
  {journal} {\bibinfo  {journal} {Nucl. Phys.}\ }\textbf {\bibinfo {volume}
  {A380}},\ \bibinfo {pages} {261} (\bibinfo {year} {1982})}\BibitemShut
  {NoStop}%
%%CITATION = NUPHA,A380,261;%%
\bibitem [{\citenamefont {Mohr}\ \emph {et~al.}(2016)\citenamefont {Mohr},
  \citenamefont {Newell},\ and\ \citenamefont {Taylor}}]{Mohr:2015ccw}%
  \BibitemOpen
  \bibfield  {author} {\bibinfo {author} {\bibfnamefont {P.~J.}\ \bibnamefont
  {Mohr}}, \bibinfo {author} {\bibfnamefont {D.~B.}\ \bibnamefont {Newell}}, \
  and\ \bibinfo {author} {\bibfnamefont {B.~N.}\ \bibnamefont {Taylor}},\
  }\href {\doibase 10.1103/RevModPhys.88.035009} {\bibfield  {journal}
  {\bibinfo  {journal} {Rev. Mod. Phys.}\ }\textbf {\bibinfo {volume} {88}},\
  \bibinfo {pages} {035009} (\bibinfo {year} {2016})},\ \Eprint
  {http://arxiv.org/abs/1507.07956} {arXiv:1507.07956 [physics.atom-ph]}
  \BibitemShut {NoStop}%
%%CITATION = ARXIV:1507.07956;%%
\bibitem [{\citenamefont {Purcell}\ and\ \citenamefont
  {Sheu}(2015)}]{Purcell:2015gtm}%
  \BibitemOpen
  \bibfield  {author} {\bibinfo {author} {\bibfnamefont {J.~E.}\ \bibnamefont
  {Purcell}}\ and\ \bibinfo {author} {\bibfnamefont {C.~G.}\ \bibnamefont
  {Sheu}},\ }\href {\doibase 10.1016/j.nds.2015.11.001} {\bibfield  {journal}
  {\bibinfo  {journal} {Nucl. Data Sheets}\ }\textbf {\bibinfo {volume}
  {130}},\ \bibinfo {pages} {1} (\bibinfo {year} {2015})}\BibitemShut {NoStop}%
%%CITATION = NDTSB,130,1;%%
\bibitem [{\citenamefont {Davis}(2005)}]{Davis:2005mb}%
  \BibitemOpen
  \bibfield  {author} {\bibinfo {author} {\bibfnamefont {D.~H.}\ \bibnamefont
  {Davis}},\ }\bibfield  {booktitle} {\emph {\bibinfo {booktitle} {{Proceedings
  on 8th International Conference on Hypernuclear and strange particle physics
  (HYP 2003): Newport News, USA, October 14-18, 2003}}},\ }\href {\doibase
  10.1016/j.nuclphysa.2005.01.002} {\bibfield  {journal} {\bibinfo  {journal}
  {Nucl. Phys.}\ }\textbf {\bibinfo {volume} {A754}},\ \bibinfo {pages} {3}
  (\bibinfo {year} {2005})}\BibitemShut {NoStop}%
%%CITATION = NUPHA,A754,3;%%
\bibitem [{\citenamefont {Bernauer}\ \emph {et~al.}(2010)\citenamefont
  {Bernauer} \emph {et~al.}}]{bernauer10}%
  \BibitemOpen
  \bibfield  {author} {\bibinfo {author} {\bibfnamefont {J.~C.}\ \bibnamefont
  {Bernauer}} \emph {et~al.} (\bibinfo {collaboration} {{A1}}),\ }\href
  {\doibase 10.1103/PhysRevLett.105.242001} {\bibfield  {journal} {\bibinfo
  {journal} {Phys. Rev. Lett.}\ }\textbf {\bibinfo {volume} {105}},\ \bibinfo
  {pages} {242001} (\bibinfo {year} {2010})}\BibitemShut {NoStop}%
\bibitem [{\citenamefont {Mrowczynski}(2017)}]{Mrowczynski:2016xqm}%
  \BibitemOpen
  \bibfield  {author} {\bibinfo {author} {\bibfnamefont {S.}~\bibnamefont
  {Mrowczynski}},\ }\href {\doibase 10.5506/APhysPolB.48.707} {\bibfield
  {journal} {\bibinfo  {journal} {Acta Phys. Polon.}\ }\textbf {\bibinfo
  {volume} {B48}},\ \bibinfo {pages} {707} (\bibinfo {year} {2017})},\ \Eprint
  {http://arxiv.org/abs/1607.02267} {arXiv:1607.02267 [nucl-th]} \BibitemShut
  {NoStop}%
%%CITATION = ARXIV:1607.02267;%%
\bibitem [{\citenamefont {Abbas}\ \emph {et~al.}(2013)\citenamefont {Abbas}
  \emph {et~al.}}]{Abbas:2013rua}%
  \BibitemOpen
  \bibfield  {author} {\bibinfo {author} {\bibfnamefont {E.}~\bibnamefont
  {Abbas}} \emph {et~al.} (\bibinfo {collaboration} {ALICE}),\ }\href {\doibase
  10.1140/epjc/s10052-013-2496-5} {\bibfield  {journal} {\bibinfo  {journal}
  {Eur. Phys. J.}\ }\textbf {\bibinfo {volume} {C73}},\ \bibinfo {pages} {2496}
  (\bibinfo {year} {2013})},\ \Eprint {http://arxiv.org/abs/1305.1562}
  {arXiv:1305.1562 [nucl-ex]} \BibitemShut {NoStop}%
%%CITATION = ARXIV:1305.1562;%%
\bibitem [{\citenamefont {Acharya}\ \emph {et~al.}(2018)\citenamefont {Acharya}
  \emph {et~al.}}]{ALICE:nucleipp2017}%
  \BibitemOpen
  \bibfield  {author} {\bibinfo {author} {\bibfnamefont {S.}~\bibnamefont
  {Acharya}} \emph {et~al.} (\bibinfo {collaboration} {ALICE}),\ }\href
  {\doibase 10.1103/PhysRevC.97.024615} {\bibfield  {journal} {\bibinfo
  {journal} {Phys. Rev.}\ }\textbf {\bibinfo {volume} {C97}},\ \bibinfo {pages}
  {024615} (\bibinfo {year} {2018})},\ \Eprint
  {http://arxiv.org/abs/1709.08522} {arXiv:1709.08522 [nucl-ex]} \BibitemShut
  {NoStop}%
%%CITATION = ARXIV:1709.08522;%%
\bibitem [{\citenamefont {Acharya}\ \emph {et~al.}(2019)\citenamefont {Acharya}
  \emph {et~al.}}]{Acharya:2019rgc}%
  \BibitemOpen
  \bibfield  {author} {\bibinfo {author} {\bibfnamefont {S.}~\bibnamefont
  {Acharya}} \emph {et~al.} (\bibinfo {collaboration} {ALICE}),\ }\href@noop {}
  {\  (\bibinfo {year} {2019})},\ \Eprint {http://arxiv.org/abs/1902.09290}
  {arXiv:1902.09290 [nucl-ex]} \BibitemShut {NoStop}%
%%CITATION = ARXIV:1902.09290;%%
\bibitem [{\citenamefont {{Anielski, Jonas}}(2015)}]{anielski-HQ14}%
  \BibitemOpen
  \bibfield  {author} {\bibinfo {author} {\bibnamefont {{Anielski, Jonas}}},\
  }\bibfield  {booktitle} {\emph {\bibinfo {booktitle} {{Proceedings, Hot
  Quarks 2014: Workshop for young scientists on the physics of
  ultrarelativistic nucleus--nucleus collisions, Andalucia, Spain}}},\
  }\href@noop {} {\bibfield  {journal} {\bibinfo  {journal} {J. Phys. Conf.
  Ser.}\ }\textbf {\bibinfo {volume} {612}},\ \bibinfo {pages} {012014}
  (\bibinfo {year} {2015})}\BibitemShut {NoStop}%
\bibitem [{\citenamefont {Adam}\ \emph
  {et~al.}(2016{\natexlab{b}})\citenamefont {Adam} \emph
  {et~al.}}]{Adam:2015yta}%
  \BibitemOpen
  \bibfield  {author} {\bibinfo {author} {\bibfnamefont {J.}~\bibnamefont
  {Adam}} \emph {et~al.} (\bibinfo {collaboration} {ALICE}),\ }\href {\doibase
  10.1016/j.physletb.2016.01.040} {\bibfield  {journal} {\bibinfo  {journal}
  {Phys. Lett.}\ }\textbf {\bibinfo {volume} {B754}},\ \bibinfo {pages} {360}
  (\bibinfo {year} {2016}{\natexlab{b}})},\ \Eprint
  {http://arxiv.org/abs/1506.08453} {arXiv:1506.08453 [nucl-ex]} \BibitemShut
  {NoStop}%
%%CITATION = ARXIV:1506.08453;%%
\bibitem [{\citenamefont {Adam}\ \emph
  {et~al.}(2016{\natexlab{c}})\citenamefont {Adam} \emph
  {et~al.}}]{ALICE:deuteronppPbPb2015}%
  \BibitemOpen
  \bibfield  {author} {\bibinfo {author} {\bibfnamefont {J.}~\bibnamefont
  {Adam}} \emph {et~al.} (\bibinfo {collaboration} {ALICE}),\ }\href {\doibase
  10.1103/PhysRevC.93.024917} {\bibfield  {journal} {\bibinfo  {journal} {Phys.
  Rev.}\ }\textbf {\bibinfo {volume} {C93}},\ \bibinfo {pages} {024917}
  (\bibinfo {year} {2016}{\natexlab{c}})},\ \Eprint
  {http://arxiv.org/abs/1506.08951} {arXiv:1506.08951 [nucl-ex]} \BibitemShut
  {NoStop}%
%%CITATION = ARXIV:1506.08951;%%
\bibitem [{\citenamefont {Shebeko}\ \emph {et~al.}(2006)\citenamefont
  {Shebeko}, \citenamefont {Papakonstantinou},\ and\ \citenamefont
  {Mavrommatis}}]{Shebeko:2006ud}%
  \BibitemOpen
  \bibfield  {author} {\bibinfo {author} {\bibfnamefont {A.}~\bibnamefont
  {Shebeko}}, \bibinfo {author} {\bibfnamefont {P.}~\bibnamefont
  {Papakonstantinou}}, \ and\ \bibinfo {author} {\bibfnamefont
  {E.}~\bibnamefont {Mavrommatis}},\ }\href {\doibase
  10.1140/epja/i2005-10247-3} {\bibfield  {journal} {\bibinfo  {journal} {Eur.
  Phys. J.}\ }\textbf {\bibinfo {volume} {A27}},\ \bibinfo {pages} {143}
  (\bibinfo {year} {2006})},\ \Eprint {http://arxiv.org/abs/nucl-th/0602056}
  {arXiv:nucl-th/0602056 [nucl-th]} \BibitemShut {NoStop}%
%%CITATION = NUCL-TH/0602056;%%
\bibitem [{\citenamefont {Zhang}\ and\ \citenamefont
  {Ko}(2018)}]{Zhang:2018euf}%
  \BibitemOpen
  \bibfield  {author} {\bibinfo {author} {\bibfnamefont {Z.}~\bibnamefont
  {Zhang}}\ and\ \bibinfo {author} {\bibfnamefont {C.~M.}\ \bibnamefont {Ko}},\
  }\href {\doibase 10.1016/j.physletb.2018.03.003} {\bibfield  {journal}
  {\bibinfo  {journal} {Phys. Lett.}\ }\textbf {\bibinfo {volume} {B780}},\
  \bibinfo {pages} {191} (\bibinfo {year} {2018})}\BibitemShut {NoStop}%
%%CITATION = PHLTA,B780,191;%%
\bibitem [{\citenamefont {Sun}\ \emph {et~al.}(2019)\citenamefont {Sun},
  \citenamefont {Ko},\ and\ \citenamefont {Doenigus}}]{Sun:2018mqq}%
  \BibitemOpen
  \bibfield  {author} {\bibinfo {author} {\bibfnamefont {K.-J.}\ \bibnamefont
  {Sun}}, \bibinfo {author} {\bibfnamefont {C.~M.}\ \bibnamefont {Ko}}, \ and\
  \bibinfo {author} {\bibfnamefont {B.}~\bibnamefont {Doenigus}},\ }\href
  {\doibase 10.1016/j.physletb.2019.03.033} {\bibfield  {journal} {\bibinfo
  {journal} {Phys. Lett.}\ }\textbf {\bibinfo {volume} {B792}},\ \bibinfo
  {pages} {132} (\bibinfo {year} {2019})},\ \Eprint
  {http://arxiv.org/abs/1812.05175} {arXiv:1812.05175 [nucl-th]} \BibitemShut
  {NoStop}%
%%CITATION = ARXIV:1812.05175;%%
\bibitem [{\citenamefont {Petran}\ \emph {et~al.}(2014)\citenamefont {Petran},
  \citenamefont {Letessier}, \citenamefont {Rafelski},\ and\ \citenamefont
  {Torrieri}}]{Petran:2013dva}%
  \BibitemOpen
  \bibfield  {author} {\bibinfo {author} {\bibfnamefont {M.}~\bibnamefont
  {Petran}}, \bibinfo {author} {\bibfnamefont {J.}~\bibnamefont {Letessier}},
  \bibinfo {author} {\bibfnamefont {J.}~\bibnamefont {Rafelski}}, \ and\
  \bibinfo {author} {\bibfnamefont {G.}~\bibnamefont {Torrieri}},\ }\href
  {\doibase 10.1016/j.cpc.2014.02.026} {\bibfield  {journal} {\bibinfo
  {journal} {Comput. Phys. Commun.}\ }\textbf {\bibinfo {volume} {185}},\
  \bibinfo {pages} {2056} (\bibinfo {year} {2014})},\ \Eprint
  {http://arxiv.org/abs/1310.5108} {arXiv:1310.5108 [hep-ph]} \BibitemShut
  {NoStop}%
%%CITATION = ARXIV:1310.5108;%%
\bibitem [{\citenamefont {Wheaton}\ and\ \citenamefont
  {Cleymans}(2009)}]{Wheaton:2004qb}%
  \BibitemOpen
  \bibfield  {author} {\bibinfo {author} {\bibfnamefont {S.}~\bibnamefont
  {Wheaton}}\ and\ \bibinfo {author} {\bibfnamefont {J.}~\bibnamefont
  {Cleymans}},\ }\href {\doibase 10.1016/j.cpc.2008.08.001} {\bibfield
  {journal} {\bibinfo  {journal} {Comput. Phys. Commun.}\ }\textbf {\bibinfo
  {volume} {180}},\ \bibinfo {pages} {84} (\bibinfo {year} {2009})},\ \Eprint
  {http://arxiv.org/abs/hep-ph/0407174} {arXiv:hep-ph/0407174 [hep-ph]}
  \BibitemShut {NoStop}%
%%CITATION = HEP-PH/0407174;%%
\bibitem [{\citenamefont {Andronic}\ \emph {et~al.}(2006)\citenamefont
  {Andronic}, \citenamefont {Braun-Munzinger},\ and\ \citenamefont
  {Stachel}}]{Andronic:2005yp}%
  \BibitemOpen
  \bibfield  {author} {\bibinfo {author} {\bibfnamefont {A.}~\bibnamefont
  {Andronic}}, \bibinfo {author} {\bibfnamefont {P.}~\bibnamefont
  {Braun-Munzinger}}, \ and\ \bibinfo {author} {\bibfnamefont {J.}~\bibnamefont
  {Stachel}},\ }\href {\doibase 10.1016/j.nuclphysa.2006.03.012} {\bibfield
  {journal} {\bibinfo  {journal} {Nucl. Phys.}\ }\textbf {\bibinfo {volume}
  {A772}},\ \bibinfo {pages} {167} (\bibinfo {year} {2006})},\ \Eprint
  {http://arxiv.org/abs/nucl-th/0511071} {arXiv:nucl-th/0511071 [nucl-th]}
  \BibitemShut {NoStop}%
%%CITATION = NUCL-TH/0511071;%%
\bibitem [{\citenamefont {Vovchenko}\ and\ \citenamefont
  {Stoecker}(2017)}]{Vovchenko:2016mwg}%
  \BibitemOpen
  \bibfield  {author} {\bibinfo {author} {\bibfnamefont {V.}~\bibnamefont
  {Vovchenko}}\ and\ \bibinfo {author} {\bibfnamefont {H.}~\bibnamefont
  {Stoecker}},\ }\bibfield  {booktitle} {\emph {\bibinfo {booktitle}
  {{Proceedings, 16th International Conference on Strangeness in Quark Matter
  (SQM 2016): Berkeley, California, United States}}},\ }\href {\doibase
  10.1088/1742-6596/779/1/012078} {\bibfield  {journal} {\bibinfo  {journal}
  {J. Phys. Conf. Ser.}\ }\textbf {\bibinfo {volume} {779}},\ \bibinfo {pages}
  {012078} (\bibinfo {year} {2017})},\ \Eprint
  {http://arxiv.org/abs/1610.02346} {arXiv:1610.02346 [nucl-th]} \BibitemShut
  {NoStop}%
%%CITATION = ARXIV:1610.02346;%%
\bibitem [{\citenamefont {Schnedermann}\ \emph {et~al.}(1993)\citenamefont
  {Schnedermann}, \citenamefont {Sollfrank},\ and\ \citenamefont
  {Heinz}}]{Schnedermann:1993ws}%
  \BibitemOpen
  \bibfield  {author} {\bibinfo {author} {\bibfnamefont {E.}~\bibnamefont
  {Schnedermann}}, \bibinfo {author} {\bibfnamefont {J.}~\bibnamefont
  {Sollfrank}}, \ and\ \bibinfo {author} {\bibfnamefont {U.~W.}\ \bibnamefont
  {Heinz}},\ }\href {\doibase 10.1103/PhysRevC.48.2462} {\bibfield  {journal}
  {\bibinfo  {journal} {Phys. Rev.}\ }\textbf {\bibinfo {volume} {C48}},\
  \bibinfo {pages} {2462} (\bibinfo {year} {1993})},\ \Eprint
  {http://arxiv.org/abs/nucl-th/9307020} {arXiv:nucl-th/9307020 [nucl-th]}
  \BibitemShut {NoStop}%
%%CITATION = NUCL-TH/9307020;%%
\bibitem [{\citenamefont {Abelev}\ \emph {et~al.}(2012)\citenamefont {Abelev}
  \emph {et~al.}}]{Abelev:2012wca}%
  \BibitemOpen
  \bibfield  {author} {\bibinfo {author} {\bibfnamefont {B.}~\bibnamefont
  {Abelev}} \emph {et~al.} (\bibinfo {collaboration} {ALICE}),\ }\href
  {\doibase 10.1103/PhysRevLett.109.252301} {\bibfield  {journal} {\bibinfo
  {journal} {Phys. Rev. Lett.}\ }\textbf {\bibinfo {volume} {109}},\ \bibinfo
  {pages} {252301} (\bibinfo {year} {2012})},\ \Eprint
  {http://arxiv.org/abs/1208.1974} {arXiv:1208.1974 [hep-ex]} \BibitemShut
  {NoStop}%
%%CITATION = ARXIV:1208.1974;%%
\bibitem [{\citenamefont {Abelev}\ \emph
  {et~al.}(2013{\natexlab{b}})\citenamefont {Abelev} \emph
  {et~al.}}]{Abelev:2013xaa}%
  \BibitemOpen
  \bibfield  {author} {\bibinfo {author} {\bibfnamefont {B.~B.}\ \bibnamefont
  {Abelev}} \emph {et~al.} (\bibinfo {collaboration} {ALICE}),\ }\href
  {\doibase 10.1103/PhysRevLett.111.222301} {\bibfield  {journal} {\bibinfo
  {journal} {Phys. Rev. Lett.}\ }\textbf {\bibinfo {volume} {111}},\ \bibinfo
  {pages} {222301} (\bibinfo {year} {2013}{\natexlab{b}})},\ \Eprint
  {http://arxiv.org/abs/1307.5530} {arXiv:1307.5530 [nucl-ex]} \BibitemShut
  {NoStop}%
%%CITATION = ARXIV:1307.5530;%%
\bibitem [{\citenamefont {Abelev}\ \emph
  {et~al.}(2013{\natexlab{c}})\citenamefont {Abelev} \emph
  {et~al.}}]{Abelev:2013qoq}%
  \BibitemOpen
  \bibfield  {author} {\bibinfo {author} {\bibfnamefont {B.}~\bibnamefont
  {Abelev}} \emph {et~al.} (\bibinfo {collaboration} {ALICE}),\ }\href
  {\doibase 10.1103/PhysRevC.88.044909} {\bibfield  {journal} {\bibinfo
  {journal} {Phys. Rev.}\ }\textbf {\bibinfo {volume} {C88}},\ \bibinfo {pages}
  {044909} (\bibinfo {year} {2013}{\natexlab{c}})},\ \Eprint
  {http://arxiv.org/abs/1301.4361} {arXiv:1301.4361 [nucl-ex]} \BibitemShut
  {NoStop}%
%%CITATION = ARXIV:1301.4361;%%
\bibitem [{Note1()}]{Note1}%
  \BibitemOpen
  \bibinfo {note} {To match the Bertsch-Pratt parameterisation ($R_{out},
  R_{side}, R_{long}$) employed by experiments, with the
  Yano-Koonin-Podgoretskii (\protect \ensuremath {R_{\perp }}, \protect
  \ensuremath {R_{\delimiter "026B30D }}, $R_{0}$) parameterisation used in the
  coalescence model, we identify $\protect \ensuremath {R_{\perp }}= R_{side}$,
  $\protect \ensuremath {R_{\delimiter "026B30D }}= R_{long}$ and then take $R
  = (\protect \ensuremath {R_{\perp }}^{2}\protect \ensuremath {R_{\delimiter
  "026B30D }})^{1/3} \approx (R_{side}^{2}R_{long})^{1/3}$.}\BibitemShut
  {Stop}%
\bibitem [{\citenamefont {Adam}\ \emph {et~al.}(2015)\citenamefont {Adam} \emph
  {et~al.}}]{Adam:2015pya}%
  \BibitemOpen
  \bibfield  {author} {\bibinfo {author} {\bibfnamefont {J.}~\bibnamefont
  {Adam}} \emph {et~al.} (\bibinfo {collaboration} {ALICE}),\ }\href {\doibase
  10.1103/PhysRevC.91.034906} {\bibfield  {journal} {\bibinfo  {journal} {Phys.
  Rev.}\ }\textbf {\bibinfo {volume} {C91}},\ \bibinfo {pages} {034906}
  (\bibinfo {year} {2015})},\ \Eprint {http://arxiv.org/abs/1502.00559}
  {arXiv:1502.00559 [nucl-ex]} \BibitemShut {NoStop}%
%%CITATION = ARXIV:1502.00559;%%
\bibitem [{\citenamefont {Aamodt}\ \emph {et~al.}(2011)\citenamefont {Aamodt}
  \emph {et~al.}}]{Aamodt:2011mr}%
  \BibitemOpen
  \bibfield  {author} {\bibinfo {author} {\bibfnamefont {K.}~\bibnamefont
  {Aamodt}} \emph {et~al.} (\bibinfo {collaboration} {ALICE}),\ }\href
  {\doibase 10.1016/j.physletb.2010.12.053} {\bibfield  {journal} {\bibinfo
  {journal} {Phys. Lett.}\ }\textbf {\bibinfo {volume} {B696}},\ \bibinfo
  {pages} {328} (\bibinfo {year} {2011})},\ \Eprint
  {http://arxiv.org/abs/1012.4035} {arXiv:1012.4035 [nucl-ex]} \BibitemShut
  {NoStop}%
%%CITATION = ARXIV:1012.4035;%%
\bibitem [{\citenamefont {Abelev}\ \emph
  {et~al.}(2013{\natexlab{d}})\citenamefont {Abelev} \emph
  {et~al.}}]{Abelev:2012sq}%
  \BibitemOpen
  \bibfield  {author} {\bibinfo {author} {\bibfnamefont {B.}~\bibnamefont
  {Abelev}} \emph {et~al.} (\bibinfo {collaboration} {ALICE}),\ }\href
  {\doibase 10.1103/PhysRevD.87.052016} {\bibfield  {journal} {\bibinfo
  {journal} {Phys. Rev.}\ }\textbf {\bibinfo {volume} {D87}},\ \bibinfo {pages}
  {052016} (\bibinfo {year} {2013}{\natexlab{d}})},\ \Eprint
  {http://arxiv.org/abs/1212.5958} {arXiv:1212.5958 [hep-ex]} \BibitemShut
  {NoStop}%
%%CITATION = ARXIV:1212.5958;%%
\bibitem [{\citenamefont {Adam}\ \emph {et~al.}(2017)\citenamefont {Adam} \emph
  {et~al.}}]{Adam:2015gka}%
  \BibitemOpen
  \bibfield  {author} {\bibinfo {author} {\bibfnamefont {J.}~\bibnamefont
  {Adam}} \emph {et~al.} (\bibinfo {collaboration} {ALICE}),\ }\href {\doibase
  10.1140/epjc/s10052-016-4571-1} {\bibfield  {journal} {\bibinfo  {journal}
  {Eur. Phys. J.}\ }\textbf {\bibinfo {volume} {C77}},\ \bibinfo {pages} {33}
  (\bibinfo {year} {2017})},\ \Eprint {http://arxiv.org/abs/1509.07541}
  {arXiv:1509.07541 [nucl-ex]} \BibitemShut {NoStop}%
%%CITATION = ARXIV:1509.07541;%%
\bibitem [{\citenamefont {Mart}\ \emph {et~al.}(1998)\citenamefont {Mart},
  \citenamefont {Tiator}, \citenamefont {Drechsel},\ and\ \citenamefont
  {Bennhold}}]{Mart:1996ay}%
  \BibitemOpen
  \bibfield  {author} {\bibinfo {author} {\bibfnamefont {T.}~\bibnamefont
  {Mart}}, \bibinfo {author} {\bibfnamefont {L.}~\bibnamefont {Tiator}},
  \bibinfo {author} {\bibfnamefont {D.}~\bibnamefont {Drechsel}}, \ and\
  \bibinfo {author} {\bibfnamefont {C.}~\bibnamefont {Bennhold}},\ }\href
  {\doibase 10.1016/S0375-9474(98)00441-2} {\bibfield  {journal} {\bibinfo
  {journal} {Nucl. Phys.}\ }\textbf {\bibinfo {volume} {A640}},\ \bibinfo
  {pages} {235} (\bibinfo {year} {1998})},\ \Eprint
  {http://arxiv.org/abs/nucl-th/9610038} {arXiv:nucl-th/9610038 [nucl-th]}
  \BibitemShut {NoStop}%
%%CITATION = NUCL-TH/9610038;%%
\bibitem [{\citenamefont {Vovchenko}\ \emph {et~al.}(2018)\citenamefont
  {Vovchenko}, \citenamefont {Doenigus},\ and\ \citenamefont
  {Stoecker}}]{Vovchenko:2018fiy}%
  \BibitemOpen
  \bibfield  {author} {\bibinfo {author} {\bibfnamefont {V.}~\bibnamefont
  {Vovchenko}}, \bibinfo {author} {\bibfnamefont {B.}~\bibnamefont {Doenigus}},
  \ and\ \bibinfo {author} {\bibfnamefont {H.}~\bibnamefont {Stoecker}},\
  }\href {\doibase 10.1016/j.physletb.2018.08.041} {\bibfield  {journal}
  {\bibinfo  {journal} {Phys. Lett.}\ }\textbf {\bibinfo {volume} {B785}},\
  \bibinfo {pages} {171} (\bibinfo {year} {2018})},\ \Eprint
  {http://arxiv.org/abs/1808.05245} {arXiv:1808.05245 [hep-ph]} \BibitemShut
  {NoStop}%
%%CITATION = ARXIV:1808.05245;%%
\bibitem [{\citenamefont {Citron}\ \emph {et~al.}(2018)\citenamefont {Citron}
  \emph {et~al.}}]{Citron:2018lsq}%
  \BibitemOpen
  \bibfield  {author} {\bibinfo {author} {\bibfnamefont {Z.}~\bibnamefont
  {Citron}} \emph {et~al.},\ }in\ \href@noop {} {\emph {\bibinfo {booktitle}
  {{HL/HE-LHC Workshop: Workshop on the Physics of HL-LHC, and Perspectives at
  HE-LHC Geneva, Switzerland, June 18-20, 2018}}}}\ (\bibinfo {year} {2018})\
  \Eprint {http://arxiv.org/abs/1812.06772} {arXiv:1812.06772 [hep-ph]}
  \BibitemShut {NoStop}%
%%CITATION = ARXIV:1812.06772;%%
\bibitem [{\citenamefont {Esposito}\ \emph {et~al.}(2015)\citenamefont
  {Esposito}, \citenamefont {Guerrieri}, \citenamefont {Maiani}, \citenamefont
  {Piccinini}, \citenamefont {Pilloni}, \citenamefont {Polosa},\ and\
  \citenamefont {Riquer}}]{Esposito:2015fsa}%
  \BibitemOpen
  \bibfield  {author} {\bibinfo {author} {\bibfnamefont {A.}~\bibnamefont
  {Esposito}}, \bibinfo {author} {\bibfnamefont {A.~L.}\ \bibnamefont
  {Guerrieri}}, \bibinfo {author} {\bibfnamefont {L.}~\bibnamefont {Maiani}},
  \bibinfo {author} {\bibfnamefont {F.}~\bibnamefont {Piccinini}}, \bibinfo
  {author} {\bibfnamefont {A.}~\bibnamefont {Pilloni}}, \bibinfo {author}
  {\bibfnamefont {A.~D.}\ \bibnamefont {Polosa}}, \ and\ \bibinfo {author}
  {\bibfnamefont {V.}~\bibnamefont {Riquer}},\ }\href {\doibase
  10.1103/PhysRevD.92.034028} {\bibfield  {journal} {\bibinfo  {journal} {Phys.
  Rev.}\ }\textbf {\bibinfo {volume} {D92}},\ \bibinfo {pages} {034028}
  (\bibinfo {year} {2015})},\ \Eprint {http://arxiv.org/abs/1508.00295}
  {arXiv:1508.00295 [hep-ph]} \BibitemShut {NoStop}%
%%CITATION = ARXIV:1508.00295;%%
\bibitem [{\citenamefont {Artoisenet}\ and\ \citenamefont
  {Braaten}(2011)}]{Artoisenet:2010uu}%
  \BibitemOpen
  \bibfield  {author} {\bibinfo {author} {\bibfnamefont {P.}~\bibnamefont
  {Artoisenet}}\ and\ \bibinfo {author} {\bibfnamefont {E.}~\bibnamefont
  {Braaten}},\ }\href {\doibase 10.1103/PhysRevD.83.014019} {\bibfield
  {journal} {\bibinfo  {journal} {Phys. Rev.}\ }\textbf {\bibinfo {volume}
  {D83}},\ \bibinfo {pages} {014019} (\bibinfo {year} {2011})},\ \Eprint
  {http://arxiv.org/abs/1007.2868} {arXiv:1007.2868 [hep-ph]} \BibitemShut
  {NoStop}%
%%CITATION = ARXIV:1007.2868;%%
\bibitem [{\citenamefont {Alcaraz}\ \emph {et~al.}(1999)\citenamefont {Alcaraz}
  \emph {et~al.}}]{Alcaraz:2000ss}%
  \BibitemOpen
  \bibfield  {author} {\bibinfo {author} {\bibfnamefont {J.}~\bibnamefont
  {Alcaraz}} \emph {et~al.} (\bibinfo {collaboration} {AMS}),\ }\href {\doibase
  10.1016/S0370-2693(99)00874-6} {\bibfield  {journal} {\bibinfo  {journal}
  {Phys. Lett.}\ }\textbf {\bibinfo {volume} {B461}},\ \bibinfo {pages} {387}
  (\bibinfo {year} {1999})},\ \Eprint {http://arxiv.org/abs/hep-ex/0002048}
  {arXiv:hep-ex/0002048 [hep-ex]} \BibitemShut {NoStop}%
%%CITATION = HEP-EX/0002048;%%
\bibitem [{\citenamefont {Poulin}\ \emph {et~al.}(2018)\citenamefont {Poulin},
  \citenamefont {Salati}, \citenamefont {Cholis}, \citenamefont
  {Kamionkowski},\ and\ \citenamefont {Silk}}]{Poulin:2018wzu}%
  \BibitemOpen
  \bibfield  {author} {\bibinfo {author} {\bibfnamefont {V.}~\bibnamefont
  {Poulin}}, \bibinfo {author} {\bibfnamefont {P.}~\bibnamefont {Salati}},
  \bibinfo {author} {\bibfnamefont {I.}~\bibnamefont {Cholis}}, \bibinfo
  {author} {\bibfnamefont {M.}~\bibnamefont {Kamionkowski}}, \ and\ \bibinfo
  {author} {\bibfnamefont {J.}~\bibnamefont {Silk}},\ }\href@noop {} {\
  (\bibinfo {year} {2018})},\ \Eprint {http://arxiv.org/abs/1808.08961}
  {arXiv:1808.08961 [astro-ph.HE]} \BibitemShut {NoStop}%
%%CITATION = ARXIV:1808.08961;%%
\bibitem [{\citenamefont {Schael}()}]{AMS100}%
  \BibitemOpen
  \bibfield  {author} {\bibinfo {author} {\bibfnamefont {S.}~\bibnamefont
  {Schael}},\ }\href@noop {} {\enquote {\bibinfo {title} {{AMS-100 - A Magnetic
  Spectrometer at Lagrange Point 2}},}\ }\bibinfo {howpublished} {{Presentation
  at AMS days La Palma, April 2018}}\BibitemShut {NoStop}%
\bibitem [{\citenamefont {Aramaki}\ \emph {et~al.}(2016)\citenamefont
  {Aramaki}, \citenamefont {Hailey}, \citenamefont {Boggs}, \citenamefont {von
  Doetinchem}, \citenamefont {Fuke}, \citenamefont {Mognet}, \citenamefont
  {Ong}, \citenamefont {Perez},\ and\ \citenamefont
  {Zweerink}}]{Aramaki:2015laa}%
  \BibitemOpen
  \bibfield  {author} {\bibinfo {author} {\bibfnamefont {T.}~\bibnamefont
  {Aramaki}}, \bibinfo {author} {\bibfnamefont {C.~J.}\ \bibnamefont {Hailey}},
  \bibinfo {author} {\bibfnamefont {S.~E.}\ \bibnamefont {Boggs}}, \bibinfo
  {author} {\bibfnamefont {P.}~\bibnamefont {von Doetinchem}}, \bibinfo
  {author} {\bibfnamefont {H.}~\bibnamefont {Fuke}}, \bibinfo {author}
  {\bibfnamefont {S.~I.}\ \bibnamefont {Mognet}}, \bibinfo {author}
  {\bibfnamefont {R.~A.}\ \bibnamefont {Ong}}, \bibinfo {author} {\bibfnamefont
  {K.}~\bibnamefont {Perez}}, \ and\ \bibinfo {author} {\bibfnamefont
  {J.}~\bibnamefont {Zweerink}} (\bibinfo {collaboration} {GAPS}),\ }\href
  {\doibase 10.1016/j.astropartphys.2015.09.001} {\bibfield  {journal}
  {\bibinfo  {journal} {Astropart. Phys.}\ }\textbf {\bibinfo {volume} {74}},\
  \bibinfo {pages} {6} (\bibinfo {year} {2016})},\ \Eprint
  {http://arxiv.org/abs/1506.02513} {arXiv:1506.02513 [astro-ph.HE]}
  \BibitemShut {NoStop}%
%%CITATION = ARXIV:1506.02513;%%
\end{thebibliography}%
\end{document}